\documentclass{aa}  

\usepackage{graphicx}
\usepackage{txfonts}
\usepackage{natbib}
\bibpunct{(}{)}{;}{a}{}{,} 

\newcommand\mone{$^{-1}$}
\newcommand\mtwo{$^{-2}$}

\newcommand\tcd{\object{3C~236}}
\newcommand\codu{\element[][12]{CO(2-1)}}
\newcommand\msun{$M_{\sun}$}
\newcommand\hd{H$_2$}
\newcommand\mhd{$M$(H$_2$)}
\newcommand\hi{\ion{H}{i}}
\newcommand\vs{$\varv_{\rm{sys}}$}
\newcommand\vsco{$\varv_{\rm{sys}}^{\rm{CO}}$}
\newcommand\msyr{\msun\ yr\mone}
\newcommand\ssfr{$\Sigma_{\rm{SFR}}$}
\newcommand\shd{$\Sigma_{\rm{MH2}}$}

\begin{document}

\title{Fueling the central engine of radio galaxies}
\subtitle{II. The footprints of AGN feedback on the ISM of \tcd. } 
\author{
A. Labiano\inst{1}
\and
S. Garc\'ia-Burillo\inst{2}
\and
F. Combes\inst{3}
\and
A. Usero\inst{2}
\and
R. Soria-Ruiz\inst{2}
\and \\
G. Tremblay\inst{4}
\and 
R. Neri\inst{5}
\and 
A. Fuente\inst{2}
\and
R. Morganti\inst{6,7}
\and
T. Oosterloo\inst{6,7}
}
\offprints{\\ Alvaro Labiano: {\tt alvaro.labiano@cab.inta-csic.es } \smallskip \\
{\it Based on observations carried out with the IRAM Plateau de Bure Interferometer. IRAM is supported by INSU/CNRS (France), MPG (Germany) and IGN (Spain). }}

\institute{
Centro de Astrobiolog\'ia (CSIC-INTA), Carretera de Ajalvir km. 4, 28850 Torrej\'on de Ardoz, Madrid, Spain.
\and
Observatorio Astron\'omico Nacional, Alfonso XII, 3, 28014, Madrid, Spain.
\and
Observatoire de Paris, LERMA \& CNRS: UMR8112, 61 Av. de l'Observatoire, 75014 Paris, France.
\and
European Southern Observatory, Karl-Schwarzschild-Str. 2, 85748 Garching bei M\"unchen, Germany.
\and
IRAM, 300 rue de la Piscine, Domaine Universitaire, 38406 St. Martin d'H\'eres Cedex, France.
\and
Netherlands Foundation for Research in Astronomy, Postbus 2, 7990 AA, Dwingeloo, The Netherlands.
\and
Kapteyn Astronomical Institute, University of Groningen, PO Box 800, 9700 AV Groningen, The Netherlands.
}

\date{  }

\abstract
{
There is growing observational evidence of
{{ active galactic nuclei (AGN) feedback on the ISM of 
 radio-quiet and radio-loud galaxies. While AGN feedback is expected 
to be more common at high redshift objects, the study of local universe galaxies help to better characterize the different manifestations of AGN feedback. 
}}
}
{
Molecular line observations 
{{can be used to quantify}}  the mass and energy budget of the gas  affected by AGN feedback.  We study the emission of molecular gas in \tcd, a 
Faranoff-Riley type 2 (FR~II) radio source at $z$$\sim$0.1, and search for the footprints of AGN feedback.  \tcd\  shows signs of a reactivation of its AGN triggered by a recent  minor merger episode.  Observations have also previously identified an extreme \hi\ outflow in this source.
} 
{
The IRAM Plateau de Bure interferometer (PdBI) has been used to study the distribution and kinematics of molecular gas in \tcd\ by imaging  with high spatial resolution (0.6$\arcsec
$) the emission of the 2--1 line of $^{12}$CO in the nucleus of the galaxy. We have searched for outflow signatures in the CO map.  {{We have also}} derived the star-formation 
rate (SFR) in \tcd\  using data {{available from}} the literature at UV, optical and IR wavelengths, to determine the star-formation efficiency  of  molecular gas. 
}
{
The CO emission in \tcd\ comes from a spatially resolved $\sim$1.4$\arcsec$(2.6~kpc)-diameter disk characterized by a regular rotating pattern.  Within the limits imposed by the sensitivity and velocity coverage of the CO data, we do not detect any outflow signatures in the cold molecular gas. 
The  disk has a cold gas mass \mhd$\sim$2.1$\times$10$^9$ \msun.  Based on CO we determine a new value for the redshift of the source $z_{\rm{CO}}$=0.09927$\pm$0.0002. The similarity between the CO and \hi\ profiles indicates that the deep \hi\  absorption in \tcd\ can be accounted for by a rotating \hi\ structure. 
This restricts the evidence of \hi\ outflow only to the most extreme velocities.  In the light of the new redshift value, the analysis of the ionized gas  
kinematics reveals a fast ($\sim$1000~km s\mone) outflow.  As for the CO emitting gas,  
outflow signatures are nevertheless absent in the warm molecular gas emission traced by infrared 
H$_2$ lines. The star-formation efficiency in \tcd\, 
 is  consistent with the value measured in $normal$ galaxies, which follow the canonical 
Kennicutt-Schmidt (KS) relation. This result, confirmed to hold in other $young$ radio sources examined in this work, is in stark contrast with the factor of 10--50 lower SFE that 
has been claimed  to characterize $evolved$ powerful radio galaxies.
}  
{
There are no signs of ongoing AGN feedback on the molecular ISM of \tcd. The recent  reactivation of the AGN in \tcd\ (about $\sim$10$^5$yr ago) is a likely explanation for the $early$ 
evolutionary status of its molecular disk. 
}
\keywords{Galaxies: individual: \tcd\ -- Galaxies: ISM -- Galaxies: kinematics and dynamics -- Galaxies: active -- ISM: jets and outflows}

\maketitle
%

\section{Introduction}



\subsection{AGN Feedback}

{Active Galactic Nuclei (AGN) release vast amounts of energy into the interstellar medium (ISM) of their host galaxies. This energy input can heat gas in the ISM, preventing its collapse and inhibiting star formation. It may also expel the gas of the ISM in the form of winds (outflows) which deplete the host of star-forming fuel. This transfer of energy from the AGN to the host is known as AGN feedback. AGN feedback could be responsible of the correlations between black hole and host galaxy bulge mass \citep{Magorrian98, Tremaine02, Marconi03, Haring04}. AGN feedback can also explain the fast transition of early type galaxies from the blue-cloud to the red-sequence \citep{Schawinski07, Kaviraj11}.
}


Over the last decade, AGN feedback has been increasingly included in models of galaxy evolution \citep{King03, Granato04, Matteo05, Croton06, Ciotti07, Menci08, King08, Merloni08, Narayanan08, Silk10}. {Observational searches for signatures of AGN feedback have also grown in number \citep{Thomas05, Murray05, Schawinski06, Muller06, Feruglio10, Crenshaw10, Fischer11, Villar11, Dasyra11, Sturm11, Maiolino12, Aalto12}.} Yet, observational evidences of AGN feedback are still fragmentary.

{AGN feedback takes place through two mechanisms: the {\it radiative} or {\it quasar} mode, and the  {\it kinetic} or {\it radio} mode. In the quasar mode, the radiation from the AGN dominates the energy transfer to the ISM. In the radio mode, the momentum of the jet is transferred to the ISM. 
}

The most dramatic effects of the kinetic mode are seen in powerful radio galaxies. With jets up to several Mpc--size, they can inject enormous amounts of energy, not only in their host galaxy ISM, but also in the IGM of the galactic group or cluster where they reside \citep[e.g.,][]{Birzan04,Fabian06,McNamara07,McNamara12}. It is thought that a significant fraction of massive galaxies undergo a radio galaxy phase at least once in their lifetime \citep[e.g.,][]{Best06}. {Consequently, understanding how AGN feedback works during this phase is crucial to a better understanding of galaxy evolvution.}


Observations of \hi\ and ionized gas in radio galaxies have found massive outflows in a significant number of radio galaxies at different redshift ranges \citep[e.g.,][]{Morganti03c, Rupke05, Holt06, Nesvadba06, Nesvadba08, Lehnert11}. Even though molecular gas may dominate the mass/energy  budget of the wind, unambiguous evidence of a molecular outflow has only been found thus far in one radio galaxy \citep[4C~12.50;][]{Dasyra11}.
Another manifestation of AGN feedback, the inhibition of star formation, has also been claimed to be at work in radio galaxies. \citet{Nesvadba10} used the 7.7 $\mu$m polycyclic aromatic hydrocarbons (PAH) emission to estimate the star-formation rate (SFR) in a sample of radio galaxies. Based on their estimated molecular gas contents,  \citet{Nesvadba10} found that radio galaxies in their sample are $\sim$10--50 times less efficient in forming stars compared to normal galaxies that follow the canonical Kennicutt-Schmidt relationship \citep{Schmidt59, Kennicutt98}. The warm-\hd\ emission of these radio galaxies suggests the presence of shocks in the warm molecular gas traced by a set of IR H$_2$ lines. These shocks, identified in the warm-H$_2$ phase,  would increase the turbulence in the molecular ISM as a whole, inhibiting the star formation in the host.
However,  it is still an open issue whether shocks are affecting the bulk of the molecular ISM or if they only concern the  warm molecular gas phase. 
CO observations, well adapted to trace most of the molecular ISM, are key if we are to understand the {complexity} of how AGN feedback affects the kinematics and the star-formation properties of molecular gas in radio galaxies.

\subsection{\tcd}


\tcd\ is the second largest radio galaxy in the Universe \citep[$\sim$4.5~Mpc deprojected size,][]{Willis74, Barthel85}\footnote{\citet{Machalski08} recently reported the discovery of the currently largest radio galaxy, J1420--0545, with a projected size of 4.7 Mpc.}. Its radio structure shows an old ($2.6\times10^8$ yr), classical double, FR II source \citep{Fanaroff74} on large scale maps, and a younger ($10^5$ yr), $\sim$2 kpc Compact Steep Spectrum source \citep[CSS, ][]{O'Dea98} in the center, responsible of two thirds of the total radio emission of \tcd\ \citep{Schilizzi01}. Both structures are roughly aligned and oriented at an angle $\sim$$30^o$ to the plane of the sky, with the North-West jet approaching \citep{Schilizzi01}. The double-double morphology \citep[CSS + large-scale source; e.g.,][]{Schoenmakers00, Kaiser00} of \tcd\ is consistent with a reignition of the AGN activity, probably due to a minor merger \citep{O'Dea01}. 

The host of \tcd\ is a massive \citep[$\sim$$10^{12-13}$ \msun,][]{Sandage72, Strom80} elliptical galaxy \citep{Barthel85, Mezcua11}, with distorted optical {morphology at kpc scales} \citep[][]{Smith89}, consistent with a merger. High resolution imaging and {optical broadband-filter absorption maps \citep{Koff00} show an inner  dust disk} (0.5 kpc from the nucleus), and a broad, dynamically young, $\sim$8 kpc long dust lane (1.5 kpc from the nucleus), misaligned by $\sim$$25\degr$ with the inner disk \citep{Martel99, Koff00}. 
The total mass of the whole dust system is $\sim$$10^7$ \msun\ \citep{Koff00, Sodroski94}.




The total stellar mass of \tcd\ has been estimated to be 10$^{12}$ \msun\ \citep{Tadhunter11}, consisting of a young ($\lesssim$$10^7$~yr) and an old stellar population ($\gtrsim$10 Gyr). The old stellar population represents 72\% of the total stellar mass \citep[see also][]{Holt07, Buttiglione09}. Using high resolution optical and UV photometry  \citet{O'Dea01} and \citet{Tremblay10} found young  ($\sim$$10^7$ yr) star forming knots on the edge of the dust lane, surrounding the nucleus, as well as and older ($\sim$$10^9$ yr) population in the nucleus. The star formation in the young knots could have been triggered by the infalling gas from the minor merger which reignited the AGN. However, it is not completely clear how the nuclear population relates to the AGN activity \citep{Tremblay10}.

The star-formation history of \tcd\ and the possibly related reignition of its AGN make \tcd\ an ideal candidate for searches of AGN feedback signatures. 
Furthermore, \tcd\ is known to have a fast ($\sim$1500 km s\mone), massive ($\sim$50 \msun\ yr\mone) \hi\ outflow: Westerbork Radio Telescope spectra show a shallow, narrow blueshfited \hi\ absorption component with a broad blue wing \citep{Morganti05}. 
Previous searches of molecular gas in \tcd\ were plagued by limited sensitivity and they were undertaken with single-dish telescopes. Based on CO observations done with the IRAM-30m telescope in a sample of radio galaxies,  \citet{Saripalli07} reported  the non-detection of CO  line emission in \tcd; the implied upper limit on the molecular gas mass of \tcd\ 
($\sim$2--3$\times$10$^9$ \msun) was not very compelling, however.




This is the second of a series of papers where we use interferometric observations of the molecular line emission to study the fueling and the feedback of activity in a sample of nearby radio galaxies \citep[observations of \object{4C~31.04}, a young compact symmetric object (CSO) source, were presented by][]{Burillo07}. 
To this purpose, we have carried out 1~mm/3~mm continuum and \codu\ line high resolution and high sensitivity observations of \tcd\ with the Plateau de Bure Interferometer (PdBI). 
We study the continuum emission of the source at 1~mm and 3~mm {wavelengths} and analyze the distribution and kinematics of molecular gas based on the emission of the \codu\ line. 
We search for outflow signatures in the CO map and re-evaluate the evidence of outflow signatures in different ISM tracers.   Furthermore, we derive the star-formation 
rate (SFR) of \tcd\  based on several tracers available {from} the literature.  Based on the gas mass derived form CO emission and on the SFR estimates, we compare the star-formation efficiency of \tcd\ against a sample of powerful radio galaxies. Finally, we discuss a scenario consistent with the properties and the history of \tcd\ and its host.

\begin{figure*}[t]
\centering
\includegraphics[width=1.5\columnwidth,angle=0]{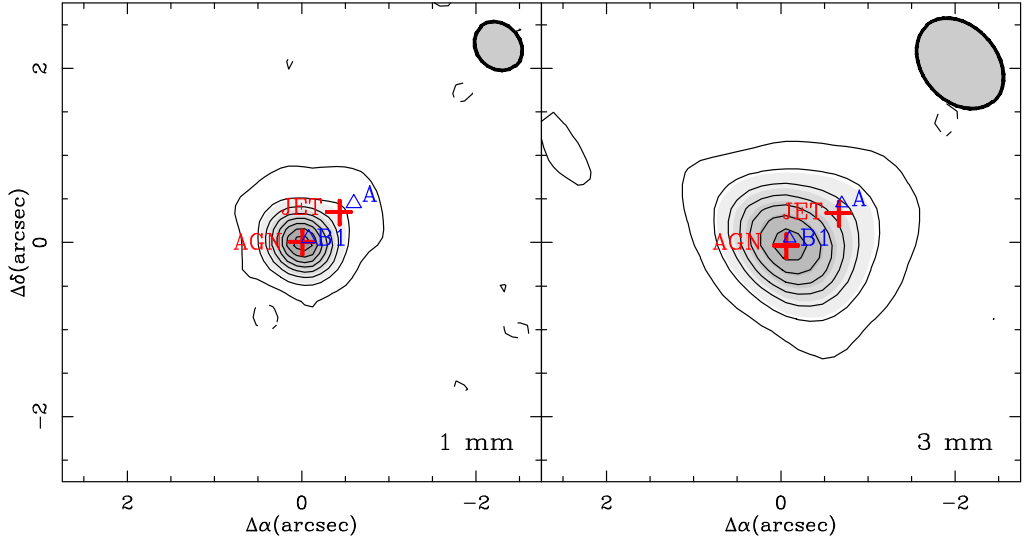} 
\caption{Continuum maps of \tcd\ at 1~mm (left) and 3~mm (right). Crosses (+) mark the position of the AGN and jet peak emission according to the UV-FIT models (Table \ref{uvfits}). 
{Triangles mark the position of components A and B1 from the VLBI map of \citet{Schilizzi01}. }
 The contour levels are --3 mJy beam\mone, 3 mJy beam\mone, 9 mJy beam\mone\ to 63 mJy beam\mone\ in steps of 9 mJy beam\mone\ for the 1~mm map, and --6 mJy beam\mone, 6 mJy beam\mone, 22~mJy beam\mone\ to 176 mJy beam\mone\ in steps of 22 mJy beam\mone\ for the 3~mm map. {The filled ellipses represent 
the beams at 1~mm (0.60$\arcsec$$\times$0.50$\arcsec$, PA=41$\degr$) and 3~mm (1.16$\arcsec$$\times$0.86$\arcsec$, PA=41$\degr$)}.
($\Delta\,\alpha$, $\Delta\,\delta$)=(0,0) corresponds to the position of the AGN at 1 mm in both panels.\label{contfits}}
\end{figure*}


\begin{table}
\caption{Point source models.}
\label{uvfits}
\begin{minipage}{1.\columnwidth}
\resizebox{1.\textwidth}{!}{
\begin{tabular}{cccccccccc} 
\hline
\hline
 Component & Wavelength & RA (J2000) &  Dec  (J2000) &    Flux (mJy)  \\ 
\hline
AGN   & 1~mm & 10 06 01.753 & +34 54 10.425 &  71.4 $\pm$ 0.4 \\ 
Jet   & 1~mm & 10 06 01.717 & +34 54 10.777 &  10.1 $\pm$ 0.4 \\ 
AGN    & 3~mm & 10 06 01.748 & +34 54 10.385 &   150.4 $\pm$ 0.8 \\ 
Jet    & 3~mm & 10 06 01.699 & +34 54 10.758 &   45.2 $\pm$ 0.8 \\ 
\hline
VLBI core  & 6 cm & 10 06 01.756 & +34 54 10.460 & --\\ 
\hline
\end{tabular}
}
\end{minipage}
\\ \\
AGN and jet positions in the 1 mm and 3 mm maps of \tcd, according to the UV-FIT models. The VLBI position of the core \citep{Schilizzi01, Taylor01} is included for comparison.
\end{table}


\section{Observations}

\subsection{Interferometer observations}


{PdBI  \citep{Guilloteau92} observations of \tcd\ were obtained in the \codu\ 1mm line using six antennas in the array B's configuration in January 2009.
}
We assumed the optical redshift  of the source derived by \citet{Holt05phd}, $z$=0.10054 ($\varv_0$(HEL)=30\,129~km~s\mone), to tune the 1~mm receivers centered on the 
$^{12}$CO(2--1) line redshifted to 209.504\,GHz.  
The relative velocity scale, although re-determined in Sect. \ref{comaps}, initially refers to this redshift ($\varv$-$\varv_o$). With this setting we obtained a velocity coverage of $\sim$1400~km\,s$^{-1}$ at 209.5\,GHz  with the narrow band correlator of the PdBI (1~GHz-wide) and the two polarizations of the receiver. Observations were conducted in a single pointing of size $\sim$23$\arcsec$. The adopted phase-tracking center of the observations was set at ($\alpha_{2000}$, $\delta_{2000}$)=(10$^{\rm h}06^{\rm m}01.7^{\rm s}$, 34$^\circ$54$\arcmin$10$\arcsec$), the position of the nucleus given by NED. Nevertheless the position of the dynamical center determined in Sect. \ref{contmap}, which  coincides with the position of the radiocontinuum VLBI core \citep{Schilizzi01}, is $\simeq$1$\arcsec$ offset to the NE with respect to the array center: ($\Delta\,\alpha$, $\Delta\,\delta$)$\sim$(0.6$\arcsec$, 0.4$\arcsec$). Visibilities were obtained through on-source integration times of  22.5 min framed by short (2 min) phase and amplitude calibrations on the nearby quasars \object{0923+392} and  \object{1040+244}. The visibilities were calibrated using the antenna-based scheme. The absolute flux density scale was calibrated on \object{MWC349} and found to be accurate to $<$10\% at 209.5\,GHz. 


{The image reconstruction was done with the standard IRAM/GILDAS software \citep{Guilloteau00}\footnote{http://www.iram.fr/IRAMFR/GILDAS}. We used natural weighting and no taper to generate the CO line map with a size of 66$\arcsec$ and 0.13$\arcsec$ /pixel sampling, 
 and obtained  a synthesized beam of 0.60$\arcsec \times$0.51$\arcsec$\,@$PA$=41$\degr$ at 1~mm.
The 1$\sigma$ point source sensitivity was derived from emission-free channels resulting in 1.1\,mJy\,beam$^{-1}$ in 20\,MHz($\sim$29\,km\,s$^{-1}$)-wide  channels.  
 We obtained a map of the continuum emission at 209.5\,GHz 
 averaging channels free of line emission from $\varv$-$\varv_o$=+100 to +640\,km\,s$^{-1}$.
The corresponding 1$\sigma$ sensitivity of continuum emission is $\sim$0.3\,mJy\,beam$^{-1}$ at 209.5\,GHz. 
 }

We subsequently observed \tcd\ in the HCO$^+$(1--0) and HCN(1--0) transitions at 3~mm 
{for six hours} on January 2010.  For these observations we tuned the receivers at a frequency intermediate between those of the  HCO$^+$(1--0) {(89.188 GHz)} and HCN(1--0) {(88.632 GHz)} transitions redshifted to 81.017\,GHz. The velocity coverage at  81\,GHz was $\sim$3700~km\,s$^{-1}$. We used the same phase tracking center and redshift as adopted for the 1~mm observations described above. The twofold goal of the 3~mm observations was the detection of the HCO$^+$(1--0) and HCN(1--0) lines as well as of their underlying continuum emission.  Notwithstanding, these lines  were not detected down to the sensitivity limit (0.8 mJy beam\mone\ in 10 MHz, $\sim$37 km s\mone\ -wide channels).  
A 3~mm continuum map was built inside the 59$\arcsec$ field-of-view using velocity channels away from any potential contribution of line emission in the signal side band of the PdBI receivers ($\varv$-$\varv_o$=-1370 to -1740\,km\,s\mone). The corresponding 1$\sigma$ 
sensitivity  is $\sim$0.3\,mJy\,beam$^{-1}$ at   81\,GHz. 
We used natural weighting to generate the maps of {the} continuum emission and obtained  a synthesized beam of 1.17$\arcsec \times$0.86$\arcsec$\,@$PA$=41$^\circ$.



\subsection{Ancillary data}

We have used the following archival HST images: ACS/SBC/F140LP (hereafter FUV image), ACS/HRC/F555W ($V$-band), STIS/NUV-MAMA/F25SRF2 \citep[NUV,][]{O'Dea01, Allen02, Tremblay10}; WFPC2/F702W ($R$-band) and NICMOS2 ($H$-band) obtained by Sparks and collaborators \citep{Koff96, McCarthy97,Madrid06, Tremblay07, Floyd08}; \hi\ data by \citet{Morganti05}; mid-IR data from \citet{Dasyra11} and \citet{Guillard12}; Sloan Digital Sky Survey (SDSS) spectra from \citet{York00}, \citet{Abazajian09}, and \citet{Buttiglione09}.
 
We use H$_0=71$, $\Omega_M=0.27 ,  \Omega_\Lambda=0.73$ \citep{Spergel03} throughout the paper. Luminosity and angular distances are D$_{L}$=452\,Mpc and D$_{A}$=374\,Mpc; the latter gives {a physical scale of} 1$\arcsec$=1.8\,kpc \citep{Wright06}.



\begin{figure}
\centering
\includegraphics[width=0.87\columnwidth]{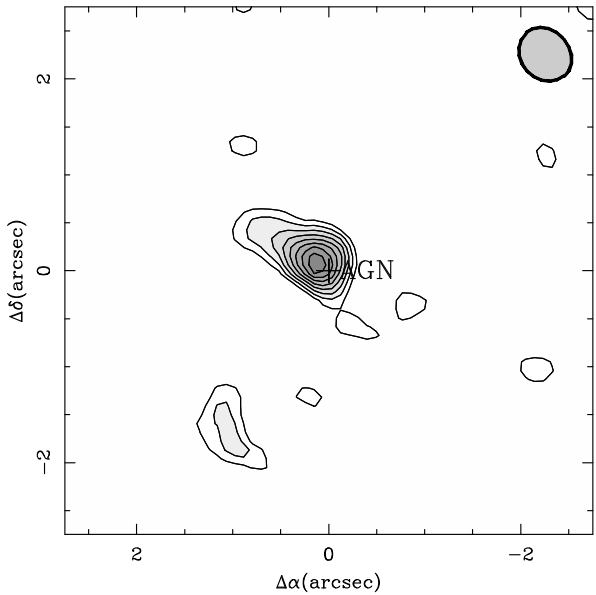} 
\caption{\codu\ intensity map of \tcd, integrating all emission above 3$\sigma$ levels, from  $\varv$-$\varv_o$=--685 to +165~km s\mone. 
The contour levels are $3\sigma$, to $6.5\sigma$ in steps of $0.5\sigma$, with $\sigma$=0.21 Jy beam\mone\ km s\mone. {($\Delta\alpha$,~$\Delta\delta$)-offsets in arcsec are relative to the location of the AGN. The gray ellipse shows the 1~mm beam.}
 \label{cointeg}}
\end{figure}

\section{Continuum maps}
\label{contmap}

Figure~\ref{contfits} shows the continuum maps derived at 209.5\,GHz and 81\,GHz in the nucleus of \tcd. 
  The emission spreads on spatial scales significantly larger than the PdBI {synthesized} beam at both frequencies and consists of a dominant central component and a {fainter} emission component that extends to the NW of the source.  
We used the GILDAS task UV-FIT to fit the continuum visibilities at both frequencies with a set of two point sources. Table~\ref{uvfits} lists the parameters  of the best fit solutions. Most of the flux comes from a central source located at ($\Delta\alpha$,~$\Delta\delta$)$\sim$(+0.6$\pm$0.1$\arcsec$,~+0.4$\pm$0.1$\arcsec$) at both frequencies. This corresponds within the errors with the position of the  AGN core determined in the VLBI 1.7 and 5~GHz maps \citep[component B2 in][]{Schilizzi01, Taylor01} and is therefore adopted as the best guess 
for the dynamical center of \tcd\  (component {labelled as} `AGN' in Fig. \ref{contfits}). The NW 1~mm and 3~mm components are both fitted by a point source located at ($\Delta\alpha$,~$\Delta\delta$)$\sim$(+0.0$\pm$0.1$\arcsec$,~+0.8$\pm$0.1$\arcsec$), i.e.,  at a position intermediate between the jet components A (radio lobe) and B1 (base of the radio jet) identified in the VLBI 1.7 and 5~GHz maps of \citet{Schilizzi01} and \citet{Taylor01}.  This indicates that the NW elongations identified in  Fig. \ref{contfits} are the mm-counterparts of the approaching radio jet of the inner  region of \tcd\  (component {labelled as} `jet' in Fig. \ref{contfits}). Using the fitted 
positions for the `AGN' and `jet' components we derive a PA of 302$^\circ$ for the mm-jet(s); this is consistent with the orientation of the CSS radio jet determined in the VLBI cm-maps of the source.

To establish the nature of the 1~mm and 3~mm emissions of \tcd\, we have compared the fluxes of the fitted components (`AGN' and `jet') with the corresponding radio emissions between 10 MHz and 10 GHz of \tcd\ (available in NED). In the log f$_\nu$ -- log $\nu$ plot, the 1~mm and 3~mm data points are aligned with the radio data points, following a power law with spectral index $\simeq$$-0.9$$\pm$0.1. This spectral index is expected for synchrotron radiation from CSS sources \citep[e.g.,][]{O'Dea98}. Therefore, we conclude that the 1~mm and 3~mm emission in \tcd\ can be accounted for by synchrotron radiation.

\begin{figure}
\centering
\includegraphics[width=0.85\columnwidth]{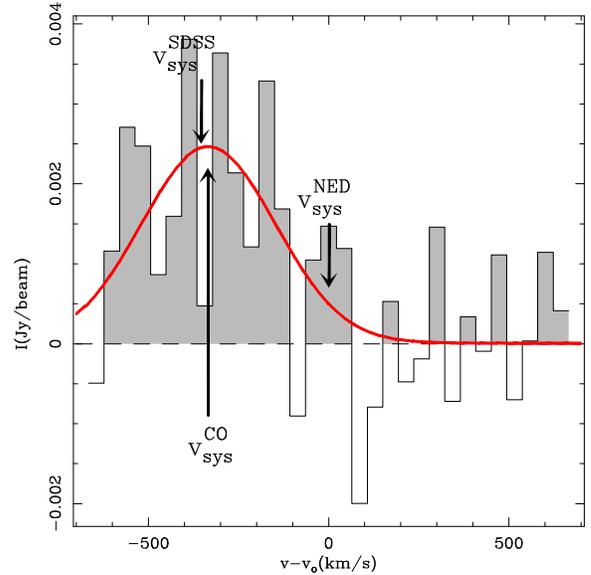} 
\caption{Spectrum of the \codu\ line emission ({after subtracting the continuum emission}) towards the AGN position of \tcd\ (histogram). The red solid line shows the Gaussian fit to the spectrum. 
Velocities in the x-axis are {with respect} to the redshift $z=0.10054$ ($\varv_0$(HEL)=30\,129~km~s\mone), where the PdBI narrow band correlator was centered.
Arrows mark the heliocentric systemic velocity of \tcd\ according to SDSS (29\,740~km~s\mone), CO (29\,761 km s\mone), and NED (30\,129 km s\mone). \label{spectrum}}
\end{figure}

\begin{figure}
\centering
\includegraphics[width=0.87\columnwidth]{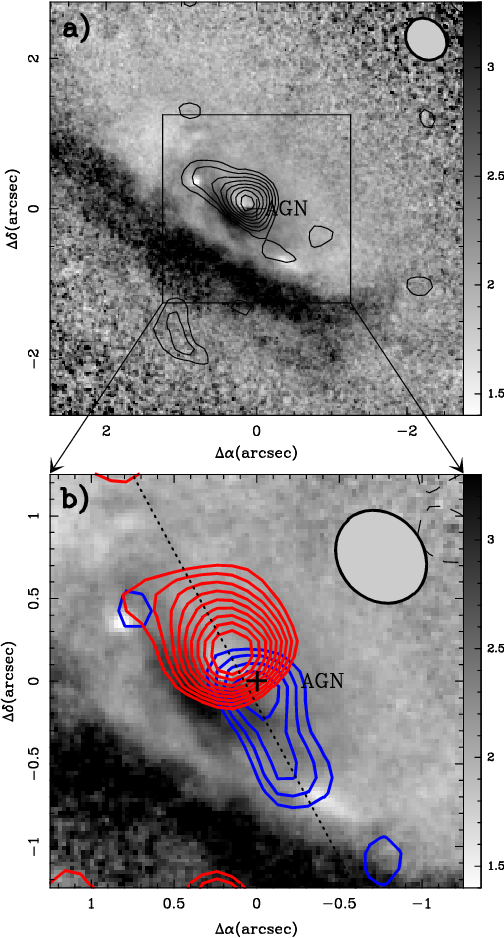} 
\caption{
{\bf a)} V--H color map (gray scale in STmag magnitude system) of \tcd, with the integrated \codu\ emission overlaid.  Contour levels as in Fig. \ref{cointeg}. 
{\bf b)} Zoom into the central 3$\arcsec$ of \tcd. Contours represent the \codu\ line maps of \tcd\ integrated in two channels: 
$blue$ (\vsco--330~km s\mone$<$$\varv$$<$\vsco) and $red$ (\vsco$<$$\varv$$<$\vsco+485~km s\mone). 
The contour levels are 3$\sigma$ to 4.5$\sigma$ in steps of 0.5 $\sigma$ ($\sigma$=0.13 Jy beam\mone\ km~s\mone) for the blue channel, and 
 3$\sigma$ to 7$\sigma$ in steps of 0.5 $\sigma$ ($\sigma$=0.13 Jy beam\mone\ km s\mone) for the red channel.
The dashed line shows the major axis of the molecular gas disk. ($\Delta\alpha$,~$\Delta\delta$)-offsets in arcsec are relative to the location of the AGN. The gray ellipses show the 1~mm beam. Color version available in electronic format. \label{vhoverlay}}
\end{figure}

\section{CO maps}
\label{comaps}
\subsection{Distribution of molecular gas}

We detect significant CO(2--1) emission above 3$\sigma$ levels from  $\varv$-$\varv_o$=--685 to +165~km s\mone. Figure \ref{cointeg} 
shows the  CO(2-1) intensity map obtained by integrating the emission within this velocity range with no threshold value 
adopted on the intensities.  As illustrated in this figure, CO emission is detected at significant levels only in the inner {diameter of} 2$\arcsec$  (3.6~kpc) of \tcd. The distribution of {the} molecular gas  can be described as a spatially-resolved elongated disk-like source. While 
the peak of CO emission lies close to the AGN, the detection of lower level emission that extends further to the NE gives the 
overall impression that the molecular disk, which has a deconvolved major axis diameter of 1.3$\arcsec$~(2.3~kpc), 
is off-centered with respect to the AGN. 

The observed asymmetry of the molecular disk can be partly attributed to the incomplete velocity coverage of the CO(2--1) 
emission in our data. As illustrated in Fig. \ref{spectrum}, which shows the CO(2--1) spectrum observed towards the AGN, any potential 
emission at velocities  $\varv$-$\varv_o$$<$--700 km s\mone\ lies outside the range covered by the PdBI narrow band correlator, 
which was initially  centered in our observations around $z=0.10054$ ($\varv_0$(HEL)=30\,129~km~s\mone). 
With this caveat in mind, the Gaussian fit to the CO spectrum of Fig.~\ref{spectrum}  provides an upper limit to the systemic velocity (\vs) 
of  the source, assuming the plausible scenario where gas motions are driven by rotation at these radii. The shape of the CO line 
profile, which shows declining intensities at the low (`blue') velocity end of the spectrum, {nevertheless indicates that}  the true 
\vs\ is {close to} the adopted value: $\varv$-$\varv_o$$<$--335 km s\mone, which corresponds to  $\varv_{\rm{sys}}^{\rm{CO}}$(HEL)=29\,761$\pm$40~km s\mone.  The implied upper limit on 
the redshift $z_{\rm{CO}}$=0.09927$\pm$0.0002 is significantly smaller  than the redshift adopted by \citet{Hill96}, who based their 
estimate on Pa$\alpha$ and H$\alpha$ measurements \citep[$z=0.10054$,][and NED]{Hill96}\footnote{\citet{Hill96} already 
noted that their redshift determination was significantly different from the previous estimate of \citet{Sandage67}, who 
measured $z=0.0989\pm0.0001$, i.e., in close agreement with the CO-based value.}. The SDSS 
has a more recent optical spectrum of \tcd\ available. 
The redshift of the source based on the SDSS spectrum is 
 $z=0.0991\pm0.0001$ (\vs(HEL)=29\,740$\pm$20 km s\mone), which is also consistent within the errors with $z_{\rm{CO}}$. In the light of the new redshift value derived from CO, we analyze in Section \ref{outf} the evidence of outflow signatures in different tracers
of the ISM of \tcd.

{Figure \ref{cointeg} shows a CO emission component SE of the AGN, unrelated to the rest of the CO. The size of this component is larger than the beam, and the flux is above the 3$\sigma$ levels, an indication that this feature {could be} real. This component falls outside the outer dust disk and beyond the radio emission of the CSS source. It is also misaligned with the radio jets. There are no features in the VLBI, optical and UV maps at the location of this component, {so its origin is unknown.} 
}

Figure \ref{vhoverlay}a shows the \codu\  map superposed on the V--H color map {made from HST imaging}. The color map shows two prominent dust 
lane features that extend over significantly different spatial scales: the {\it outer} dust lane visualizes a highly-inclined gas disk of 
$\sim$4.5$\arcsec$ (8.1~kpc)--major axis diameter oriented along $PA$$\simeq$$55^\circ$. The {\it inner} dust lane pictures a much smaller disk-like structure of $\sim$1$\arcsec$ (1.8~kpc)--major axis diameter oriented along $PA$$\simeq$$30$$^\circ$. The two concentric disks are 
{thus misaligned}, a characteristic already noted by \citet{O'Dea01} and interpreted as a signature of {discrete} accretion events. The $inner$ disk 
is roughly perpendicular to the inner radio jet, oriented along $PA$$\simeq$$302^\circ$.

\subsection{Kinematics of molecular gas}

As shown in Fig.~4a, the CO disk seems to be closely linked to the $inner$ dusty disk: to the limit of our sensitivity, CO emission is not detected in the 
$outer$ disk. To better illustrate this association, we show in Fig.~4b the CO emission integrated in two channels defined to cover the ranges 
corresponding to $blue$ velocities (\vsco--330~km s\mone$<$$\varv$$<$\vsco) and $red$ velocities 
(\vsco$<$$\varv$$<$\vsco+485~km s\mone). The velocity structure of the CO disk is spatially resolved in the maps: the emission peaks 
in the $red$ and $blue$ lobes are separated by $\simeq$0.45$\arcsec$, i.e., a significant fraction of the PdBI beam. Lower-level yet significant emission, 
which stands out more clearly in Fig.~4b than in Fig.~4a thanks to the narrower velocity channels used to generate the maps, extends up to radial 
distances $\sim$0.7$\arcsec$ (1.3~kpc) from the AGN.   The line joining the peaks of the $blue$ and $red$ lobes has a $PA$=28.07$^\circ$, which is consistent with the orientation of the $inner$ 
dusty disk. 
 
\begin{figure}
\centering
\includegraphics[width=0.85\columnwidth]{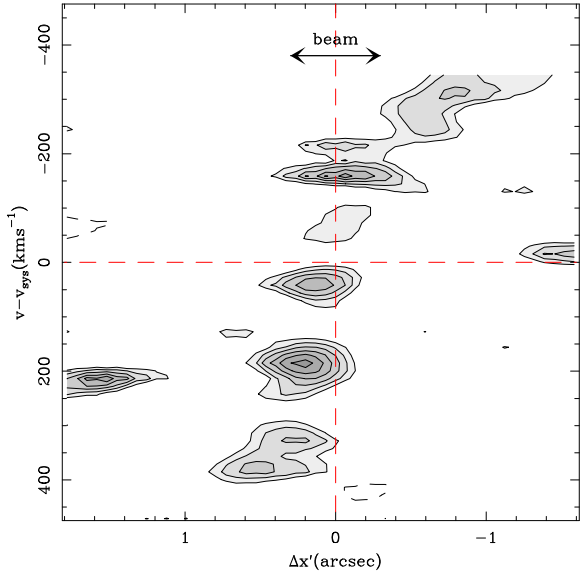} 
\caption{Position-velocity diagram 
of the \codu\ emission along the major axis of the molecular gas disk of \tcd\ (dashed line in Fig. \ref{vhoverlay}). Positions ($\Delta$x) are relative to the AGN. Velocities are relative to the CO \vsco. The contour levels are --2$\sigma$, 2$\sigma$ to 4.5$\sigma$ in steps of 0.5$\sigma$, with $\sigma$=3~mJy~beam\mone\ km s\mone. 
\label{pvdiagram}}
\end{figure}

Figure \ref{pvdiagram} shows the position-velocity (p-v) diagram taken along this line ($PA$=28.07$^\circ$), identified as the kinematical major axis of the CO disk. The kinematical pattern of the 
p-v diagram {reveals} the signature of a spatially-resolved rotating molecular gas disk. At the NW edge of the CO disk, the radial velocity reaches $\sim$400~km s\mone. If we assume that the 
inclination of the gas disk is 60$^\circ$,  based on the orientation of the radio source 
\citep{Schilizzi01}, we derive that $\varv_{\rm{rot}}$$\sim$460~km~s\mone\ at $R$$\sim$1.2~kpc.  
 {This is similar to the typical range of rotation velocities derived in  other early-type galaxies \citep[FWZI$\lesssim$800 km s\mone; e.g.,][and references therein]{Krips10,Crocker12}.
 }

Based on a spherical mass distribution model, the dynamical mass $M_{\rm{dyn}}$ inside $R$, derived as  $M_{\rm{dyn}}$=$R$$\times$$\varv_{\rm{rot}}^2$/G (where G is the gravitational constant,  R is the de-projected radius of the disk, and $\varv_{\rm{rot}}$ is the de-projected radial velocity at the edge of the disk), is 
$\sim$6.1$\times10^{10}$ \msun. Accounting for the mass of molecular gas derived in Sect~\ref{mass}, we determine that
the spheroidal stellar mass ($M_{\rm{sph}}$) is  $\sim$5.8$\times10^{10}$ \msun\  inside  $R$$\sim$1.2~kpc if the contribution from dark matter to M$_{\rm{dyn}}$ is neglected. Not surprisingly, this 
value falls short of accounting for the total stellar mass $M_*$ of \tcd, which amounts to $9.2\times 10^{11}$ \msun\ according to the stellar population models of \citet{Tadhunter11}.
This value, together with the available estimates of the supermassive black hole mass ($M_{\rm{bh}}$) can be used to evaluate if \tcd\ fits within the known $M_{\rm{bh}}$-$M_{\rm{sph}}$ relation followed 
by different galaxy populations. The two estimates for  $M_{\rm{bh}}$) in \tcd\ published by \citet{Marchesini04} ($M_{\rm{bh}}$=4.2$\times$10$^8$ \msun) and \citet{Mezcua11} 
($M_{\rm{bh}}$=3.2$\times$10$^8$ \msun) are in rough agreement.  Taking the average of both estimates, the implied $M_{\rm{bh}}$/$M_{\rm{sph}}$ ratio in \tcd\ is  $\sim$0.05$\%$. This is compatible 
within the errors with the $M_{\rm{bh}}$/$M_{\rm{sph}}$ ratio predicted for radio-loud AGNs at $z$$\sim$0.1, based on to the redshift-dependent law found by  \citet{McLure06} (see Eq.~3 of their paper).


 

\subsection{Mass of molecular gas}\label{mass}

The total flux obtained within the 21$\arcsec$   primary beam field of the PdBI is  4~Jy~km s\mone.  Assuming a brightness temperature ratio CO(2--1)/CO(1--0)$\simeq$1, we can use the equations 
of \citet{Solomon97} to derive the equivalent CO(1--0) luminosity: $L$$\arcmin_{\rm{CO}}=4.6\times10^8$ K km s\mone\ pc\mtwo. If we apply the Galactic ratio of \hd-mass to CO-luminosity \citep[4.6 \msun/K 
km s\mone\ pc$^2$,][]{Solomon87}, we estimate that the H$_2$ mass in \tcd\ is \mhd=2.1$\times$$10^9$ \msun. Including the mass of helium, the corresponding total molecular gas mass is $2.9\times10^9$ \msun. The average column density of hydrogen atoms derived from CO is $N_{\rm{H,CO}}=9.4$$\times$$10^{22}$ cm$^{-2}$.
{
The {\it Chandra} spectrum of \tcd, which has  a spatial resolution ($\sim$$0.6\arcsec$) similar to our observations,  is best fitted by a power-law plus an absorption component with  $N_{\rm{H,CO}}=2$$\times$$10^{22}$ cm$^{-2}$ (Birkinshaw et al., 2012 in prep.). The column densities derived from CO and X-rays are thus consistent within a factor 4--5.
}






Strong emission from mid-IR H$_2$ rotational lines, a tracer of warm ($T_{\rm{k}}$$>$100~K) and diffuse ($n$(H$_2)$$\sim$10$^{3}$cm$^{-3}$) molecular hydrogen, has also been detected in \tcd. 
The galaxy qualifies as a MOlecular Hydrogen Emission Galaxy (MOHEG) {due to} its large H$_2$ to PAH(7.7$\mu$m) luminosity ratio $L$(H$_2$)/$L$(7.7$\mu$m)$\geq$0.33, i.e., well 
beyond the limit defined by \citet{Ogle10}: $L$(H$_2$)/$L$(7.7$\mu$m)$>$0.04.  \citet{Dasyra11} used {\it Spitzer} IRS to study the warm-\hd\ emission lines ((0-0) S0 28.22 $\mu$m, (0-0) S1 
17.04 $\mu$m, (0-0) S2 12.28 $\mu$m, (0-0) S3 9.66 $\mu$m) of \tcd. Based on the flux of the S1, S2 and S3 lines,  \citet{Dasyra11} computed that the total mass of warm-\hd\ in \tcd\ is 6.10$\times10^7$ \msun, 
{with a single excitation temperature of $T$=345~K. }
Using the same data set, \citet{Guillard12} fitted a LTE model of three excitation temperature 
components to the emission of the warm-\hd\ in \tcd. According to their model, the warm-\hd\ phase is composed of ($1.6\pm0.5$)$\times10^9$ \msun\  of  gas at $T$=100~K, plus 
($9.5\pm4.0$)$\times10^7$~\msun\ at $T$=241~K and $7.3\pm1.8\times10^5$~\msun\ for $T$=1045~K. The warm-\hd\ phase is more likely to have a distribution of temperatures, rather than one fixed 
temperature. Hence, we adopt the \citet{Guillard12} model in our estimate of the warm-to-cold \hd\ ratio in \tcd, for which we obtain a value of 
0.81.  
Gas-rich star forming galaxies usually present a warm-to-cold \hd\ ratio of 0.01--0.1 \citep[e.g.,][]{Higdon06, Roussel07}. \citet{Ogle10} show that this ratio is significantly higher in powerful 
radio galaxies classified as MOHEGs, where the ratio ranges from $\sim$0.2 to $\sim$2.  The  ratio measured in \tcd\ lies within the values found in MOHEGs.

Although the optical line ratios measured in \tcd\ are consistent with those of low excitation radio AGNs \citep[LERAGN,][]{Smolcic09, Hardcastle06},  the \mhd/$M_*$ ratio ($\sim$0.23$\%$) and the  $M_{\rm{bh}}$/$M_{\rm{sph}}$ ratio ($\sim$0.05$\%$) measured in  \tcd\ are similar to the values found in high excitation radio AGNs \citep[HERAGN,][]{Smolcic11}. The mixed
properties of \tcd\ could be a consequence of the recent merger-driven reactivation of star formation and AGN  activities in this source\footnote{LERAGN are usually hosted by galaxies on their final stages of the mass assembly, with older stellar populations and redder colors compared to their HERAGN counterparts.}.

\begin{figure}[t]
\centering
\includegraphics[width=\columnwidth]{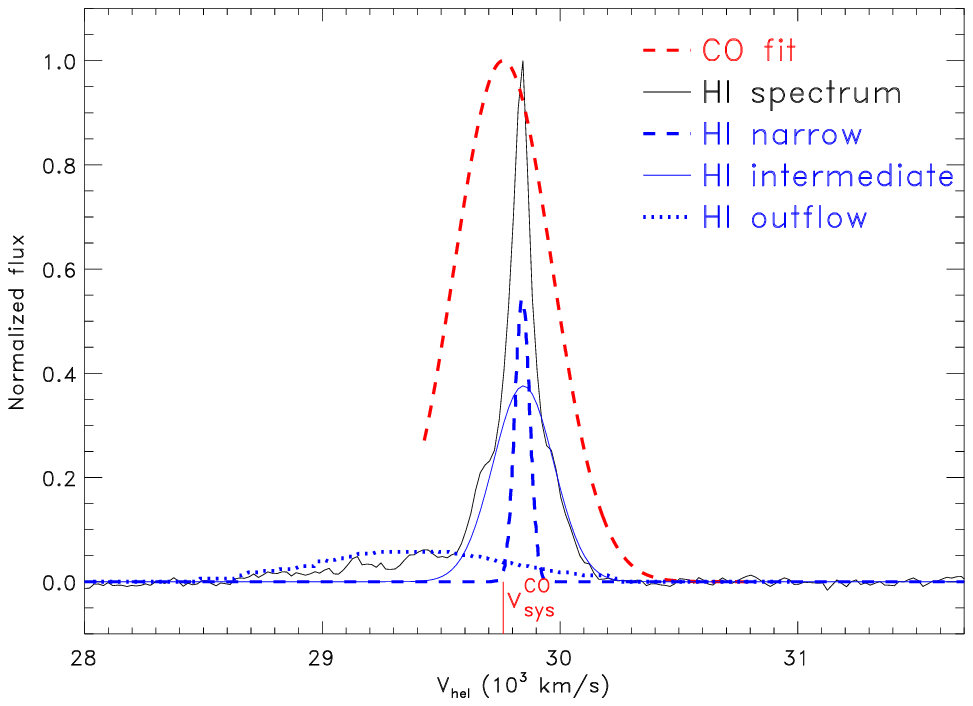} 
\includegraphics[width=\columnwidth]{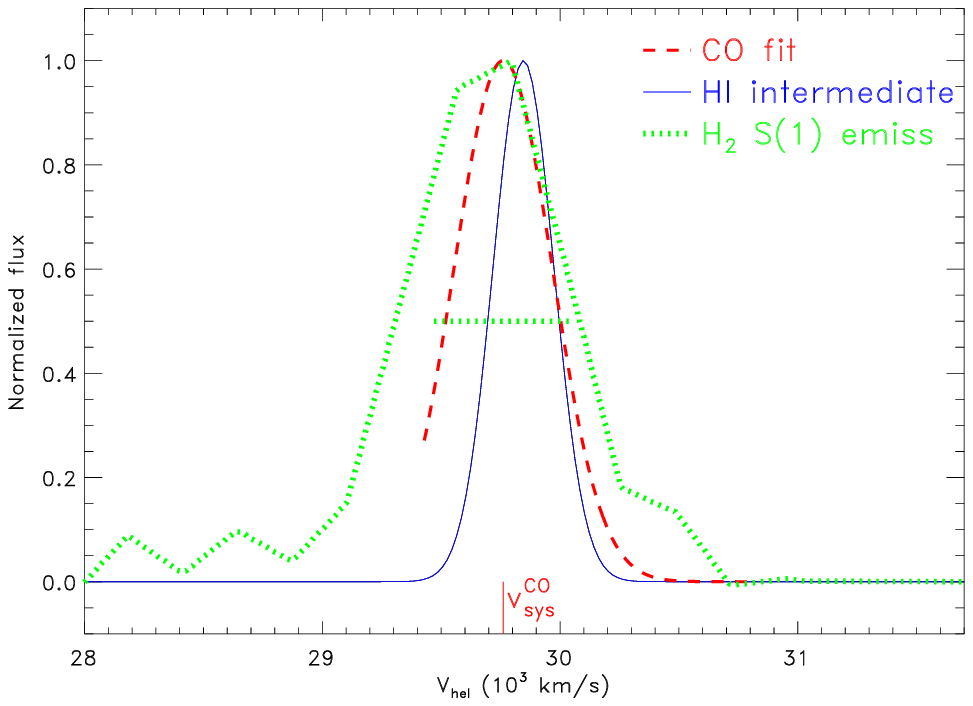} 
\caption{
\hi\ absorption \citep{Morganti05}, \hd\ S(1) 17 $\mu$m emission  \citep{Dasyra11}, and Gaussian fit of the \codu\ emission of \tcd.  For comparison, the spectra have been normalized, and the \hi\ absorption is shown inverted. The red vertical line marks the \vsco.
Top panel: Comparison of the CO emission (red, dashed line) and \hi\ absorption. The black line shows the integrated \hi\ absorption spectrum of \tcd. The three components of the fit to the \hi\ spectrum are represented with blue lines.
Bottom panel:  Comparison of the CO emission (red, dashed line), \hd\ S(1) 17 $\mu$m emission (green line), and the \hi\ intermediate width component (blue line). {The horizontal green line shows the FWHM of the \hd\ S(1) 17 $\mu$m line, corrected for instrumental broadening.} \label{cohuno}}
\end{figure}
\begin{figure}[t]
\centering
\includegraphics[width=0.75\columnwidth,angle=-90]{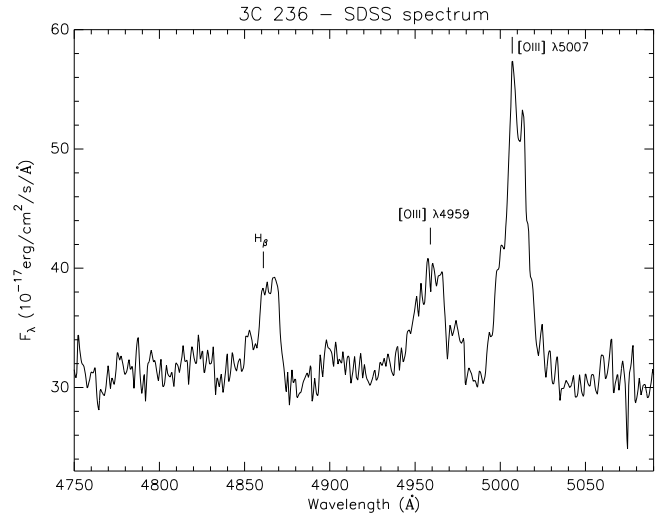} 
\caption{Optical rest-frame spectrum of \tcd\ from SDSS ($z_{\rm{SDSS}}$=0.0991). For clarity, we only show the region of the H$\beta$ and [\ion{O}{iii}]$\lambda\lambda$4959,5007 \AA\ emission lines. 
 \label{sdssoutflowuno}}
\end{figure}

\begin{figure}[t]
\centering
\includegraphics[width=0.9\columnwidth,angle=0]{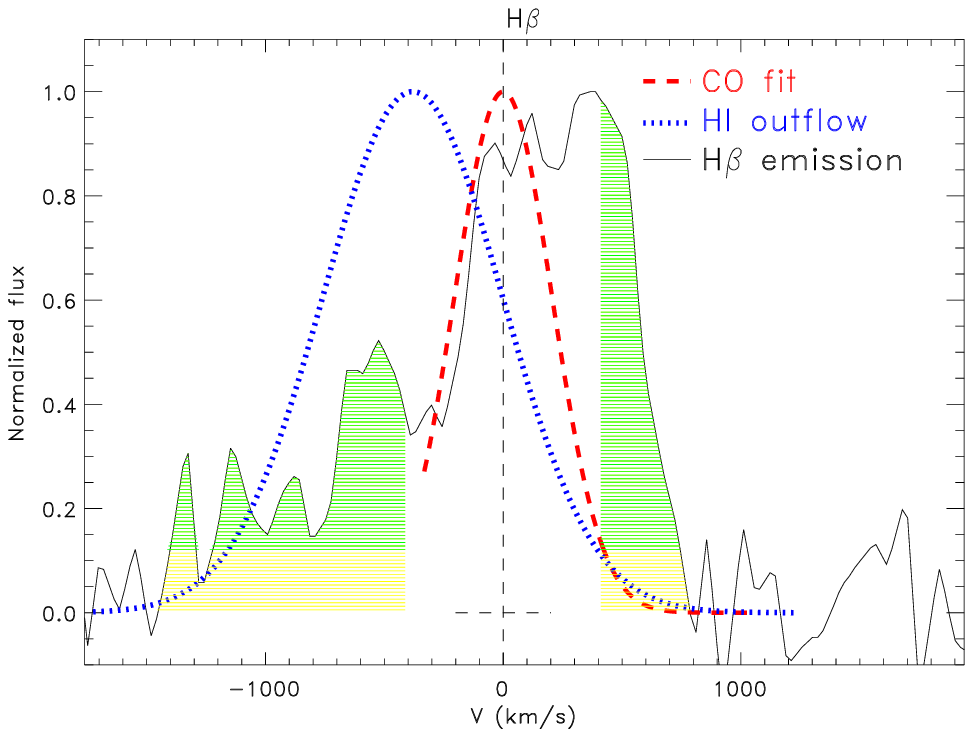}
\includegraphics[width=0.9\columnwidth,angle=0]{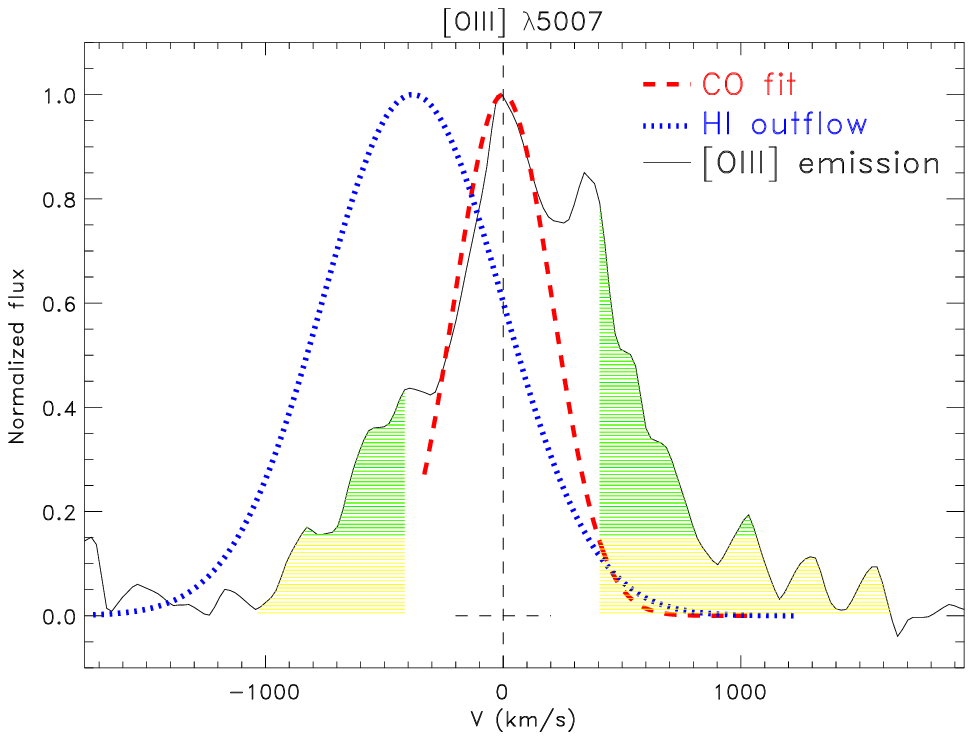} 
\caption{
Zooms of the  H$\beta$ (top panel) 
 and [\ion{O}{iii}]$\lambda$5007 \AA\ (bottom panel) emission lines (black, solid line), with the \codu\ emission (red, dashed line), and \hi\ outflow absorption (blue, dotted line) overlaid. For comparison, all fluxes have been normalized, and the \hi\ absorption has been inverted.
The horizontal dashed line shows the FWHM of \codu\ emission. 
The vertical dashed line marks the \vsco. 
{The shaded areas mark the emission of the optical lines beyond $\pm$400 km s\mone, above the continuum (yellow) and 3$\sigma$ (green) levels.}
 Color version available in electronic format. \label{sdssoutflowdos}}
\end{figure}

\section{Evidence of outflow in \tcd}
\label{outf}


Based on the HI spectrum of \tcd\ obtained with the WSRT, which shows a 2000~km~s$^{-1}$--wide absorption profile, \citet{Morganti05} identified in this source an extreme  outflow of atomic hydrogen.  
In the light of the new redshift of the source determined from CO,  we discuss below the existence of outflow signatures that can be found from the analysis of the kinematics of different ISM components of \tcd.

\subsection{Cold molecular gas: CO line emission}

{The kinematics of the spatially-resolved CO disk, analyzed in Sect.~\ref{comaps}, can be explained by circular rotation around the AGN.  Our data do not sample the  extreme blueshifted end of the \hi\ outflow (beyond $\varv$-\vsco$<$$-332$ km s\mone). We nevertheless note that  the data show no signs of significant emission 
at the high end represented by extreme $red$ velocities (883 km s\mone{$>$}$\varv$-\vsco$>$$368$ km s\mone), which are fully covered in our observations. The implied upper limit to the molecular mass of the $red$ lobe of the outflow is $\lesssim$4.4$\times10^8$ \msun. 
We used the properties of the molecular outflow of \object{Mrk~231} as an extreme upper limit to what could be expected in \tcd:
  if we adopt  in \tcd\ the same scaling for the outflow/disk mass ratio measured in \object{Mrk~231} \citep[i.e., \mhd$_{\rm{outflow}}$/\mhd$_{\rm{disk}}$$\sim$0.06,][]{Feruglio10,Cicone12}, the expected \hd\  mass in the outflow would amount to 1.2$\times10^8$ \msun. This is a still factor $\simeq$4 below the detection limit derived above for \tcd.  We can therefore conclude that the presence of a molecular  outflow that would share the extreme properties measured in \object{Mrk~231} would  have remained unnoticed in \tcd\ by our observations.}

\subsection{Atomic gas: HI line absorption}

Figure \ref{cohuno} shows the integrated \hi\ absorption profile of \tcd\ \citep{Morganti05}. The spectrum has been inverted for the sake of comparison with the CO emission profile.  The \hi\ {line} profile shows three distinct velocity components hereafter referred to as {\it narrow}, {\it intermediate}, and {\it outflow} components. The {\it narrow} and {\it intermediate} components were already detected by \citet{Gorkom89} using the VLA. The {\it outflow} component has been confirmed with EVLA observations (A array, Morganti {\it et al.} in prep.). The velocity centroids and widths derived from the Gaussian fits  to the {\it narrow} and {\it intermediate} components ($narrow$: $\varv$(HEL)=29\,828 km s\mone, FWHM$\sim$80 km s\mone; $intermediate$: $\varv$(HEL)=29\,846 km s\mone, FWHM$\sim$300 km s\mone) differ significantly from the parameters of the {\it outflow} component ($\varv$(HEL)=29\,474 km s\mone, FWHM$\sim$1100 km s\mone).  As argued below, the \hi\ spectrum can be explained by two kinematically distinct systems: a rotating disk and an outflow. 

The main limitation of \citet{Morganti05} data, if we are to pinpoint the origin of the different components of the \hi\ spectrum, is their lack of spatial resolution.  The comparison between the integrated \hi\ and CO profiles can be used to discuss the possible location of the \hi\ absorbers, however. As shown in Fig.~\ref{cohuno},  the velocity range covered by the $narrow$ and $intermediate$ components is similar to the one covered by CO emission.  This similarity can be taken as indirect evidence that the deep  \hi\  absorption is explained by a rotating \hi\ structure that is likely concomitant with the molecular disk. This would leave out the evidence of \hi\ outflow only to the most extreme velocities (--1000~km s\mone$<$$\varv-$\vsco$<$--500~km s\mone)\footnote{We note that previous claims of an extreme \hi\ outflow in \tcd\ were mainly  due to the wrong value assumed for \vs.}. The new VLBI map of \tcd\ presented by \citep{Struve12} has recently confirmed this scenario by showing that the \hi\ absorption of the $narrow$ and $intermediate$ component  come from a rotating structure characterized by an orientation almost identical to that of the CO disk.  The location of the \hi\ $outflow$ remains unknown, however, because  the velocity range of the $outflow$ lies well beyond the bandwidth  covered by  the data of \citet{Struve12}.

\subsection{Warm molecular gas:  \hd\ line emission}

 
 
 Figure \ref{cohuno} (bottom panel) shows the \hd\ S(1) emission line profile  detected in \tcd\ by \citet{Dasyra11}\footnote{The S(1), S(2) and S(3) lines detected in \tcd\ show similar FWHM \citep{Dasyra11}.}. A comparison of the S(1) line with the Gaussian fit to the   \codu\  spectrum and with the \hi\ $intermediate$ component indicates that the line profiles of these tracers are similar (Fig.~\ref{cohuno}, bottom panel). The velocity centroids and deconvolved widths of the lines are almost identical within the errors ($\Delta$$\varv$$\sim$50 and 60 km s\mone\ for \hi\ and \hd\ centroids respectively). Even though the apparent FWHM of the S1 line  (750$\pm$75 km s\mone) is $\sim$50$\%$ larger compared to the CO line and the \hi\ $intermediate$ component, this difference disappears if the instrumental broadening of the S(1) line is taken into account (FWHM=582$\pm$60~km~s\mone, corrected for instrumental broadening). This similarity can be taken as indirect evidence that the warm-\hd\ emission of \tcd\ is probably generated in the same rotating disk as the CO emission.






\subsection{Ionized gas: optical line emission}
\label{iongasoutf}

{Figure \ref{sdssoutflowuno} shows the optical emission lines H$_\beta$ and [\ion{O}{iii}] $\lambda\lambda$4959,5007 \AA\  of \tcd, taken from the SDSS archive.  
The three lines show a remarkable large width, also seen in the optical emission of H$\alpha$, [\ion{N}{ii}]$\lambda\lambda$6548,6584 \AA, [\ion{S}{ii}]$\lambda\lambda$6713,6731 \AA\ lines, as well as in the IR lines [\ion{Ne}{ii}] and [\ion{Ne}{iii}]  \citep{Dasyra11, Guillard12}. Figure \ref{sdssoutflowdos} shows a zoom on the H$_\beta$ and [\ion{O}{iii}] $\lambda$5007 \AA\ lines, with  
the Gaussian fit to the \codu\  spectrum and the \hi\ $outflow$ component overlaid. 
This figure  shows that a sizable fraction of the total emission of the lines (above 3$\sigma$-levels) is detected at extreme `red' and `blue' velocities: $\mid$$\varv-$\vsco$\mid$$>$400~km s\mone, i.e.,  well beyond the range allowed by the rotation of the disk (see discussion in Sect.~\ref{comaps}). While the velocity centroids of the optical lines are similar to \vsco, their widths (FWHM$\sim$900 km s\mone, FWZI$\sim$2000 km s\mone) are $\sim$3 times the value measured for the \codu\ line. 
 
The imprint of the outflow is clearly detected on the red and the blue wings of the optical lines, suggesting that, unlike the emission from the star-forming disk,  the emission from the outflowing gas is not as heavily extincted (see discussion in Sect.~\ref{photom}). 
The similar FWZI of the line wings suggests that the red wing is produced in the receding side of the same outflow system.
The lack of spatial resolution prevents us from clearly identifying 
the origin of the outflow, however.  A comparison with the \hi\ outflow component shows that the blue wing of the ionized gas emission lines covers a comparable velocity range, which suggests that the ionized gas outflow may have started to recombine and form \hi\ \citep[e.g.,][]{Morganti03}.

}

\section{Star-formation properties of \tcd}
\label{secsfe}


\citet{Nesvadba10}  studied the relation between the SFR surface density, and the molecular gas surface density derived from CO in a sample of radio galaxies.  The SFR was estimated based on the 7.7 $\mu$m PAH feature emission, except for \object{3C~326~N}, where the 70 $\mu$m continuum emission was also used as a SFR tracer \citep{Ogle07}.
 With all the caveats discussed by \citet{Nesvadba10} in mind, an inspection of the Kennicut-Schmidt (KS) law shown in Fig.~9 of {their paper indicates that powerful radio sources as systematically associated with PAH intensities that are a factor $\sim$10--50 lower than normal galaxies for a given mass.  }
  This offset is suggestive of a lower   star-formation efficiency ($SFE$=$SFR$/\mhd) in MOHEGs compared to  normal galaxies.

 
\citet{Nesvadba10} pointed to the influence of large-scale shocks as the agent responsible for increasing the turbulence of molecular gas \citep[see also][]{Nesvadba11}. This energy injection  in the medium could inhibit star formation to a large extent.
In Sect~6.2, we study the SFE and the location of \tcd\ in the KS diagram, and discuss the differences with respect to the results obtained by \citet{Nesvadba10} in their sample of MOHEGs. 
To this aim we have used several tracers of the SFR available for \tcd\ (discussed in Sect~6.1). This is required if we are to constrain the potential biases inherent to the different SFR calibrations.  Some SFR tracers are sensitive to the presence of AGN radiation and jet-induced shocks in the ISM, as these may increase the flux of ionized gas emission lines, vary the shape {of the continuum}, and destroy ISM molecules like the PAH. A possible evolutionary scenario explaining the differences between the samples is {discussed} in Sect.~6.3.



\begin{figure}
\centering
\includegraphics[width=\columnwidth]{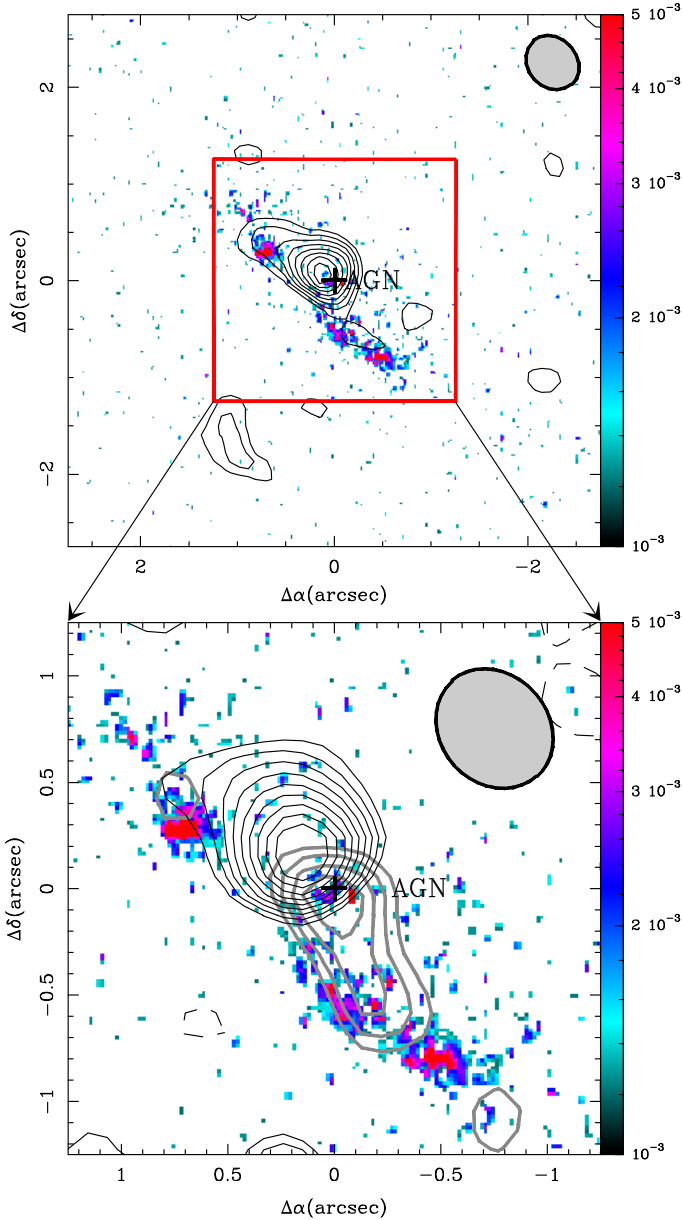} 
\caption{Top panel: HST FUV image of the star forming regions of \tcd, with the integrated \codu\ emission overlaid. Bottom panel: zoom into the central 3$\arcsec$ of \tcd. The contours correspond to the \codu\ emission integrated in the red (thin, black line) and blue (thick, gray line) channels. 
Contour levels as in Fig. \ref{vhoverlay}. Color version available in electronic format.  HST image in counts s\mone. \label{nuvoverlay}}
\end{figure}

\subsection{Star-formation rate estimations}
\label{sfrestim}


\subsubsection{Optical/Ultraviolet photometry}
\label{photom}

\citet{Tremblay10} and \citet{O'Dea01} used HST high resolution imaging in the FUV, NUV, U, and V bands to study the star formation in \tcd. As illustrated in Fig.~\ref{nuvoverlay}, the FUV emission consists of five 
knots distributed in the inner 4.5 kpc of \tcd: there is one central knot located at the AGN and four knots arranged in an arc along the outer edge of the inner dusty disk described in Sect. \ref{comaps}. 
\citet{Tremblay10} concluded that all the knots are associated with star-formation episodes yet of significantly different ages:  $\sim$$10^9$ yr for the nucleus, and $\sim$$10^7$ yr for the arc. 
 Using Starburst99 stellar population synthesis models to fit the data, \citet{Tremblay10} determined that the UV-derived SFR of  \tcd\ is 7.51 \msyr.

{Figure \ref{nuvoverlay} shows that the star forming regions of \tcd\ lie  on the nearest edge of the CO and inner dust disk, while the rest of the galaxy shows no UV emission (except for the compact nucleus). This morphology, strongly asymmetrical with
respect to the galaxy major axis,  suggests that there is a strong extinction in the inner dust disk. 
 If we calculate the extinction using the $N(H)$ values derived from CO and X-ray data (Sect.~\ref{mass}), the \citet{Bohlin78} relationships yield A$_V$=10 and A$_V$=50 respectively. 
 Therefore, the UV images of \tcd\ are probably missing UV emission from the star forming regions embedded in the inner disk and, thus, the SFR estimate of \citet{Tremblay10} is probably a lower limit of the total SFR of \tcd. The emission detected in the UV knots observed would arise from the outermost, low-extinction, layers of the inner dust disk.

The extinction of the inner dust disk does not seem to significantly absorb the emission of the ionized gas outflow, which shows {both} red and blue wings in the spectrum (Sect. \ref{iongasoutf}). Based on the emission line fluxes from the SDSS spectrum of \tcd, the \citet{Cardelli89} extinction law yields significantly lower extinction values: A$_V$=1.4 \citep[using $E(B-V)$=0.456,][]{Tremblay10}. 
This suggests that the outflow is formed in a different region of \tcd, probably the compact nucleus, which is visible in the UV images. 
High-resolution spectroscopy would be nevertheless needed to identify the location of the outflow.   

}


UV and optical emission of powerful radio galaxies may have a significant contribution from the AGN. This may cause a severe overestimation of the  SFR if this contribution is not deducted. 
\citet{Tremblay10} discussed that the contribution from emission lines to their UV and optical bands is negligible for \tcd, however.  A similar conclusion was reached by
\citet{Holt07}, who fitted the  shape of the continuum emission subjacent to the optical spectrum of \tcd\ inside the range 3000--7000 \AA. In their analysis,  \citet{Holt07} found no evidence of a significant AGN component. 

Some CSS sources are known to show UV nebulosities that are roughly co-spatial with the region occupied by the expanding radio jet \citep{Labiano08}. This association suggests that both
emission features are related in these sources. In the case of \tcd\, however, the radio jets are perpendicular to the star-forming disk revealed by the HST images. This different geometry indicates that
the UV and optical knots are mostly unrelated to the jet lobes in this source.
We can therefore conclude that the UV and optical emission from the star-forming regions of \tcd\  has a negligible contribution from the AGN.  The value derived for the SFR (7.51 \msyr) from these tracers is in all likelihood not significantly biased.

\subsubsection{Optical spectroscopy}
\label{halfa}

The H$\alpha$ emission of a galaxy is indicative of its SFR \citep{Kennicutt98}. 
Using the H$\alpha$ line flux from SDSS \citep[$F_{\mathrm{H}\alpha} = 5.62\times10^{-15}$ erg s\mone\ cm\mtwo,][]{Buttiglione09},  and the H$\alpha$-SFR relationships of \citet{Kennicutt98}, we obtain a $SFR$=1.1 \msun\ yr\mone\ for \tcd.
If a galaxy harbors an AGN, the nuclear activity can alter the H$\alpha$ flux through photoionization and/or shocks. 
To estimate the contributions from the AGN to the H$\alpha$ emission, we use the so called diagnostic diagrams or Baldwin-Phillips-Terlevich (BPT) diagrams \citep{Baldwin81, Veilleux87}. 
\citet{Buttiglione09} measured the de-reddened ratios of the optical emission lines in the SDSS spectrum of \tcd. 
A comparison of the values of [\ion{O}{iii}]/H$\beta$, [\ion{N}{ii}]/H$\alpha$, [\ion{S}{ii}]/H$\alpha$ and  [\ion{O}{i}]/H$\alpha$ from \citet{Buttiglione09} with the results of \citet{Kewley01} and \citet{Brinchmann04} shows that the contribution of the AGN to the H$\alpha$ flux of \tcd\ is $\lesssim$10\%. Removing 10\% of the flux of the H$\alpha$ line yields $SFR$=1.0~\msun~yr.   Based on the same BPT diagrams, we find that the optical line ratios of \tcd\ are also compatible with ionization from shocks with velocities $\lesssim$300 km s\mone. These shocks could be generated either by the AGN or by star~formation processes.

The  4000 $\AA$ break \citep[D4000,][]{Balogh99} is indicative of the SFR per unit of stellar mass ($M_*$) in AGN \citep{Brinchmann04}. For \tcd, D4000$\simeq$1.8 \citep{Holt07}, which yields $SFR$/$M_*$$\sim$1$\times$$10^{-11}$--$10^{-12}$ yr\mone. Using the total stellar mass of \tcd\ estimated by \citet{Tadhunter11},  $M_*$=$10^{12}$ \msun, we obtain $SFR$$\sim$1--10 \msun\ yr\mone, consistent with the SFR estimations  of Sect.~\ref{photom}.

\subsubsection{Mid-infrared continuum}


\citet{Calzetti07} presented two SFR calibrations for their sample of nearby galaxies, one using 24 $\mu$m luminosity of the source, the other one combining the 24 $\mu$m and H$\alpha$ luminosities \citep[see also][]{Kennicutt09}. Applying these calibrations to the 24 $\mu$m  emission \citep[$F_{24\mu\mathrm{m}}$=2.16$\times$$10^{-12}$ erg s\mone\ cm\mtwo,][]{Dicken10}, and the H$\alpha$ emission (Sect. \ref{halfa}) of \tcd, we obtain $SFR$=6.62 \msun\ yr\mone\ and $SFR$=9.63$\pm$1.7 \msun\ yr\mone\ respectively.

\citet{Dicken10} also measured the 70 $\mu$m flux of \tcd\ with {\it Spitzer}: $F_{70\mu\mathrm{m}}$=2.77$\times$$10^{-12}$ erg s\mone\ cm\mtwo. Based on the 70  $\mu$m emission, we can establish a low and an upper limit \citep[Dicken 2011, private communication; see also][]{Seymour11} on the SFR of \tcd: 3 \msyr$\lesssim$$SFR$$\lesssim$10 \msun\ yr\mone.   

The 24 $\mu$m and 70 $\mu$m emission can be increased by the heating of dust by the AGN \citep{Tadhunter07}, producing an overestimated SFR. However,  \citet{Dicken10} found that the 70 $\mu$m emission of \tcd\ is $\sim$9 times larger than expected for a non-starburst galaxy with the same [\ion{O}{iii}] luminosity. They argue that this difference is attributable to star formation being the main contributor to the 70 $\mu$m flux. The SFR derived from the 24~$\mu$m and 70 $\mu$m are consistent with the SFR derived in Sect.~\ref{photom}, suggesting 
that the contribution from the AGN to these fluxes is small, and the SFR estimations from the IR continuum are accurate.








\subsubsection{Mid-infrared spectroscopy}

\citet{Willett10} used the [\ion{Ne}{iii}]$\lambda$15.6 $\mu$m and [\ion{Ne}{ii}]$\lambda$12.8 $\mu$m fluxes to measure the SFR in a sample of Compact Symmetric Objects \citep[see also][]{Ho07}. The Neon emission lines of \tcd\ have fluxes $F_{[\ion{Ne}{ii}]}$ = (0.92$\pm$0.06)$\times$$10^{10}$ erg s\mone\ cm\mtwo\  and $F_{[\ion{Ne}{iii}]}$ = (0.44$\pm$0.03)$\times$$10^{10}$~erg~s\mone~cm\mtwo\ \citep[{\it Spitzer} spectrum,][]{Guillard12}, which yield $SFR$ = 16$\pm$1~\msun~yr\mone.

To assess the AGN contribution to the total MIR luminosity of a galaxy, \citet{Willett10} compare the [\ion{O}{iv}]$\lambda$25.9 $\mu$m / [\ion{Ne}{ii}] $\lambda$12.8 $\mu$m ratio with the equivalent width of the 6.2 $\mu$m PAH emission. We measured [\ion{O}{iv}]/[\ion{Ne}{ii}]$<$0.2 and, $EW_{6.2}$$<$170 $\mu$m (3$\sigma$ limit) from the {\it Spitzer} spectrum of \tcd. Hence, the AGN contribution in \tcd\ is $<$10\%. \citet{Genzel98} used the 7.7 $\mu$m PAH strength, instead of the 6.2 $\mu$m, to study the contribution of the AGN to the galaxies in their sample. If we consider the 7.7 $\mu$m PAH strength of \tcd, the contribution from the AGN is $\lesssim$20\%, consistent with the samples of {\it star-formation dominated} galaxies \citep{Genzel98}. Assuming that the AGN contributes in the same proportion to the MIR luminosity and the Neon lines, removing $20\%$ of the Neon lines flux yields $SFR$=12.9$\pm$0.5 \msun\ yr\mone. 

Another indication of the AGN contribution is given by the ratio [\ion{Ne}{iii}]/[\ion{Ne}{ii}], which increases with the hardness of the ionizing environment. For \tcd, log [\ion{Ne}{iii}]/[\ion{Ne}{ii}]=--0.3, similar to the  mean value of the ratio in ULIRG and starburst galaxies (--0.35), and below the mean value for AGN (--0.07) and CSO \citep[--0.16,][]{Willett10}, suggesting a low contribution from the AGN in \tcd. 

\citet{Pereira10} published a list of IR emission line ratios for their sample of 426 active and HII galaxies. Based on their results, the [\ion{O}{iv}]/[\ion{Ne}{ii}] and [\ion{Ne}{iii}]/[\ion{Ne}{ii}] ratios of \tcd\ are closer to the values observed in LINER-like and HII galaxies rather than in Seyferts or QSO.  Therefore, the SFR of \tcd, obtained with IR line ratios,  is not largely overestimated due to AGN effects. It is also consistent with the SFR values derived  in Sect.~\ref{photom}. 

\subsubsection{PAH emission}
\label{sfrpah}

The {\it Spitzer} spectrum of \tcd\ shows that the emission of the PAH features is $F_{11\mu\mathrm{m}}$=3.7$\times$$10^{-14}$ erg s\mone\ cm\mtwo, $F_{7.7\mu\mathrm{m}}$$<$6.8$\times$$10^{-14}$ erg s\mone\ cm\mtwo, and $F_{6.3\mu\mathrm{m}}$$<$3.8$\times$$10^{-14}$ erg s\mone\ cm\mtwo\ \citep[3$\sigma$ limits,][]{Dicken11, Guillard12}.

Based on the 7.7 $\mu$m PAH emission limit, we used the \citet{Calzetti07} equations \citep[see also,][]{Nesvadba10} to establish a 3$\sigma$-limit on the SFR of \tcd: $SFR$$\lesssim$0.23 \msun\ yr\mone. This upper limit falls below any other SFR estimation for \tcd, and is consistent with the SFR (which was also calculated using the 7.7 $\mu$m emission) of the \citet{Nesvadba10} radio galaxies  with \hd\ masses $\sim$$10^9$ \msun. 

\citet{Willett10} used a second method to calculate the SFR in their sample, based on the correlation found between the Neon emission line luminosities and the 6.2 $\mu$m plus 11.3 $\mu$m PAH luminosities \citep[see also][]{Farrah07}. For \tcd, the total flux of the 6.2 $\mu$m and 11.3 $\mu$m PAH emission is  3.7$\times$$10^{-14}$ erg s\mone\ cm\mtwo\ $<$$F_{6.2\mu\mathrm{m}}$+$F_{11.3\mu\mathrm{m}}$$<$7.5$\times$$10^{-14}$ erg s\mone\ cm\mtwo, yielding 11 \msyr\ $\lesssim$$SFR$$\lesssim$22 \msyr. 
Leaving aside the estimate derived from the 7.7 $\mu$m PAH feature, we conclude that  the average value of the SFR obtained from  all the different tracers discussed above is $SFR$$\sim$9.2 \msun\ yr\mone (the corresponding value from the 7.7 $\mu$m emission is thirty-five  times lower). It is thus tempting to speculate that 7.7 $\mu$m emission underestimates the SFR in \tcd. 
 \citet{O'Dowd09} 
 found that sources with an AGN component have weaker 7.7 $\mu$m emission than quiescent galaxies, consistent with the destruction of smaller PAH by shocks and/or radiation from the AGN. 
Modelization of the PAH interactions with the ISM show that PAH are destroyed by shocks with velocities $\gtrsim$125 km~s\mone\ \citep[][and references therein]{Micelotta10b, Micelotta10a}. 
The  shock velocities measured in the  warm-\hd\ phase of \tcd\ are however too small \citep[$\lesssim$30 km~s\mone,][]{Guillard12} to destroy the PAH\footnote{
PAH emission is likely produced in the PDR-like phase of molecular gas at temperatures $\geq$100 K,  which is also the major contributor to the bulk of H$_2$ line emission. 
}. Another mechanism of PAH destruction is by the AGN radiation field. Hard-UV and X-rays will destroy PAH at kiloparsec distances from the AGN \citep[e.g.,][]{Siebenmorgen04, Voit92}. 
As discussed above, the star forming knots in \tcd\ are not found to be contaminated by the effects of AGN radiation, however. Similarly, the PAH destruction from hard-UV and X-rays should in all likelihood affect only the nuclear region, leaving aside the star forming knots far from the AGN. 
The reasons behind the discrepancy between  the SFR value derived from the 7.7 $\mu$m PAH and the rest of the tracers in \tcd\ remain therefore unknown.



\begin{table}[t]
\caption{SFR estimations of \tcd.}
\label{tab_sfr}
\begin{minipage}{\columnwidth}
\centering
\resizebox{\textwidth}{!}{
\begin{tabular}{cccccccc} 
\hline
\hline
Region & Tracer &  SFR    & \mhd\  & Area  &  Refs. \\                 
       &        &   (\msun\ yr\mone) & (10$^8$ \msun)& (kpc$^2$) &\\
\hline
All      & Optical+UV  & 7.51            &       21.3 & 12.7     & 1 \\ 
Unresolved & H$\alpha$ & 1.0            &     21.3 & 12.7       & 2,3 \\                          
Unresolved &  D4000    & 1-10            &     21.3 & 12.7       & 4,5 \\                          
Unresolved & 24 $\mu$m   & 6.62          &     21.3 & 12.7       & 6,7 \\                              
Unresolved & 24$\mu$m+H$\alpha$ & 9.63$\pm1.7$  &
                                                21.3 & 12.7  & 2,6,7 \\                          
Unresolved & 70 $\mu$m   & 3-10  &    21.3 & 12.7  & 6,8   \\ 
Unresolved & 7.7 $\mu$m  &  $<$0.24     &    21.3 & 12.7  & 7,9  \\                           
Unresolved & 6.2+11.3 $\mu$m  &  11-22   &    21.3 & 12.7  & 10,11 \\                            
Unresolved & [\ion{Ne}{ii}]+[\ion{Ne}{iii}] &  12.9$\pm$0.5  &    21.3 & 12.7  & 10,11 \\                            
Region 0 &     FUV     & 6.19            &       21.3 & 12.7     & \ref{appregions} \\                            
Region 1 &     FUV     & 0.84            &    $<$9.93 & 3.0      & \ref{appregions} \\                            
Region 2 &     FUV     & 1.43            &       5.49 & 2.4      & \ref{appregions} \\                            
Region 3 &     FUV     & 1.27            &       4.93 & 2.5      & \ref{appregions} \\                            
Region 4 &     FUV     & 1.48            &       3.66 & 2.5      & \ref{appregions} \\                            
Region 5 &     FUV     & 1.17            &    $<$12.5 & 2.3      & \ref{appregions} \\                            
\hline
\end{tabular}
}
\end{minipage}
\\ \\
Limits are 3$\sigma$. 
 Uncertainties in \mhd\ are $\sim$$5\%$. SFR uncertainties listed when available.\\
References:
1-\citet{Tremblay10}, 2-\citet{Buttiglione09}, 3-\citet{Kennicutt98}, 4-\citet{Holt07}, 5-\citet{Brinchmann04}, 6-\citet{Dicken10}, 7-\citet{Calzetti07}, 8-\citet{Seymour11}, 9-\citet{Guillard12}, 10-\citet{Dicken11}, 11-\citet{Willett10}, \ref{appregions}-Appendix \ref{appregions} of this work.
\end{table}

\begin{figure}
\centering
\includegraphics[width=\columnwidth,angle=0]{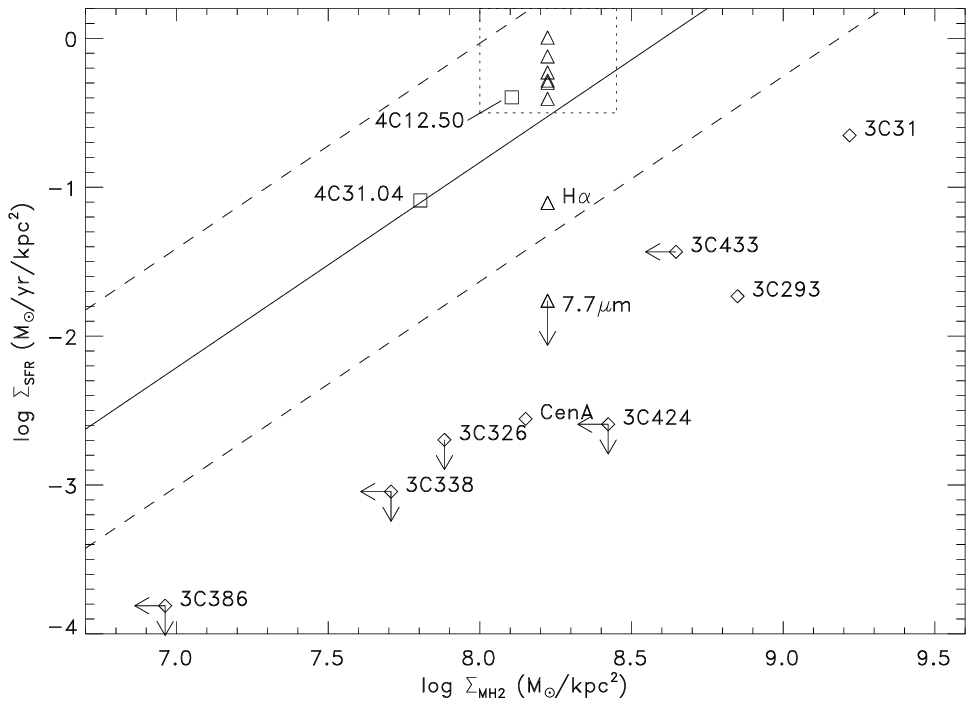} 
\includegraphics[width=\columnwidth,angle=0]{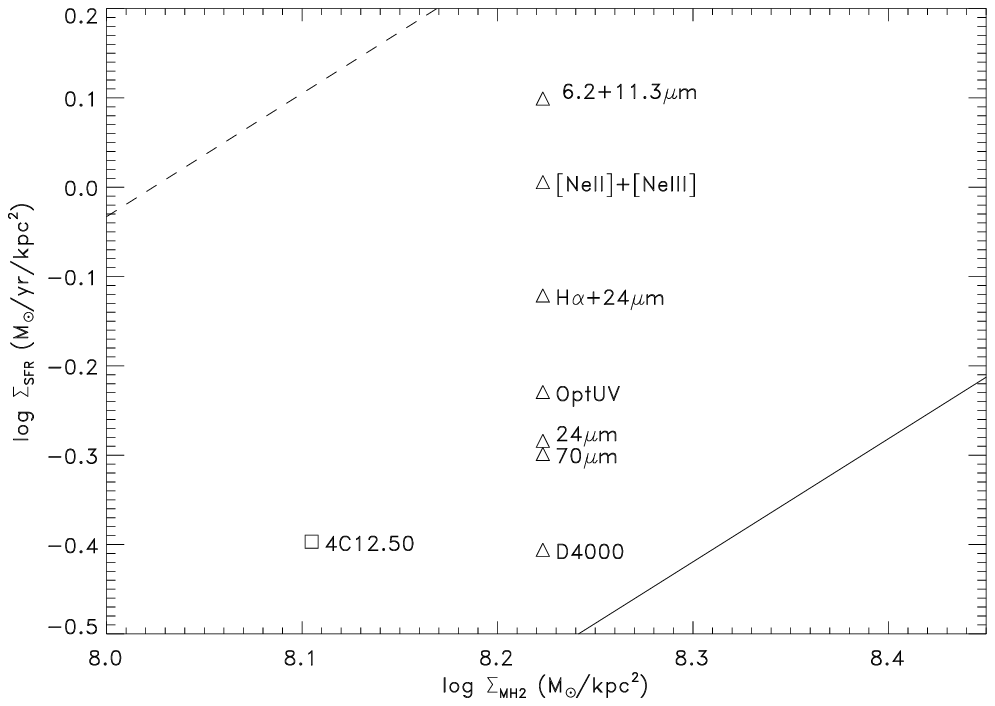} 
\caption{\ssfr\ and \shd\ of \tcd\ (triangles), the \citet{Nesvadba10} sample (diamonds), and young sources  \object{4C~12.50} and \object{4C~31.04} (squares). 
The bottom panel is a zoom on the area marked with a dotted box, where most of the SFR estimations of \tcd\ fall. 
Solid line: best-fit of the KS-law from \citet{Kennicutt98}. Dashed lines: dispersion around the KS-law best fit for normal star-forming galaxies \citep{Roussel07, Kennicutt98}. 
\label{nicoleplot}}
\end{figure}

\subsection{The star-formation efficiency in \tcd: `normal' or `inhibited'?}

Figure \ref{nicoleplot} shows the SFR surface density (\ssfr=$SFR$/$area$) and cold-\hd\ mass  surface density (\shd=$M(H_2)$/$area$) for \tcd\ and the \citet{Nesvadba10} sample\footnote{ The data available from \citet{Nesvadba10} do not include warm-\hd. Therefore, we considered only the cold-\hd\ mass for the comparison of star-formation laws, which yields log \shd$\simeq$8.2 \msun~kpc\mtwo. Adding the cold and warm-\hd\ masses of \tcd\ yields log \shd$\simeq$8.5~\msun~kpc\mtwo.}. It also shows the canonical KS-law fitted for normal star-forming galaxies \citep{Roussel07, Kennicutt98}. For the surface density calculations, we have assumed that all the  star formation of \tcd\ is produced in a disk of radius $R$=1.1$\arcsec$ (deprojected area 12.7~kpc$^2$), which includes the molecular gas disk, and all the UV emission from the star-forming knots. 
Table \ref{tab_sfr} lists all the SFR estimations for \tcd, the areas considered, and the corresponding cold-\hd\ mass.

Figure \ref{nicoleplot} shows that the \ssfr\ of \tcd\ is consistent with the \ssfr\ of normal star-forming galaxies at similar \shd. In terms of SFE, \tcd\ shows $SFE_{\rm{3C236}}$=4~Gyr\mone. 
For a normal galaxy with the same \shd\ as \tcd, the KS-law yields \ssfr=0.33 \msyr\ kpc\mtwo. Therefore, the implied $SFE_{\rm{normal}}$=2 Gyr\mone. For the \citet{Nesvadba10} sample, the SFE (0.02-0.04 Gyr\mone) is 10--50 times lower than measured in normal galaxies.
Compared to the powerful radio galaxies of \citet{Nesvadba10}, \tcd\ is   an efficient star-forming radio galaxy. 
At the current rate, \tcd\ will deplete its \hd\ gas in $t_{\rm{deplete}}$$\sim$2 Gyr.
 Our SFR calculations do not consider the extincted star-forming regions embedded in the inner dust disk (Sect.~\ref{photom}). The inclusion of these regions in the SFR estimations would increase the SFE value of \tcd\ and thus the difference with the radio galaxies of \citet{Nesvadba10}.


The HST images and CO map of \tcd\ provide enough resolution to study the spatial distribution of the SFE. In Appendix \ref{appregions}, we discuss how the \ssfr\ and \shd\ change along the molecular gas disk of \tcd. Our analysis shows that all the star-forming regions of \tcd\ have a SFE consistent with the KS-law of normal galaxies. \tcd\ shows a high global SFE compared with other radio galaxies. 
This property cannot be attributed to any particular location in the disk.

\subsection{AGN feedback in radio-galaxies: an evolutionary path?}

\tcd\ has undergone two  epochs of AGN activity, with a $\sim$10$^7$~yr phase of inactivity in-between. After this dormant phase, an accretion event (e.g., a minor merger\footnote{\citet{O'Dea01} detect a small companion galaxy at 10$\arcsec$ from the nucleus of \tcd.}  or re-settlement of the gas) increased the amount of gas, reactivated the star formation and triggered the ($\sim$10$^5$ yr old) CSS source \citep{Tremblay10, O'Dea01}. 
With the AGN inactive at the time of the accretion event, the lack of large-scale radio jets, and the lobes disconnected to the core, the feedback effects of the old radio source could not affect the newly acquired gas.
If an outflow was created at the time of the CSS ignition, it would have needed to travel at an average velocity of $20\,000$ km s\mone\ to reach the star-forming knots seeing in the UV images (located at $\sim$2 kpc from the nucleus) in 10$^5$ years. This speed is $\sim$10--20 times higher than the fastest outflows observed in radio galaxies. Then, such scenario can be excluded. The molecular gas in the knots has thus not been affected by the CSS source feedback. Therefore, the knots form stars at a normal rate. Within 10$^6$--10$^7$ yr, the feedback effects may nevertheless reach the knots and  inhibit the star formation, lowering the SFE to the values measured for the \citet{Nesvadba10} radio sources (which have ages of 10$^7$--10$^8$ years).

In the scenario described above, where the effects of AGN feedback are closely related to the age of the radio source, we may expect to measure SFE values typical of normal star-forming galaxies in the hosts of other young radio sources.
\citet{Willett10} measured the SFR of 9 young 
sources using the 6.2 and 11.3 $\mu$m PAH emission (as shown in Sect. \ref{sfrpah} for \tcd) and obtained values from 0.8 to 47.8 \msun\ yr\mone. Two of these have published cold \hd\ mass estimations: \object{4C~12.50} \citep[$SFR$=31.5 \msun\ yr\mone, \mhd=(1.0$\pm$0.3)$\times$$10^{10}$ \msun,][]{Dasyra12} and \object{4C~31.04} \citep[$SFR$=6.4 \msun\ yr\mone, \mhd=0.5$\times$$10^{10}$ \msun,][]{Burillo07}. Figure \ref{nicoleplot} shows that the SFE of both sources are comparable to the SFE of \tcd\ and thus consistent with the KS-law of normal galaxies. Although the data are scarce, young radio galaxies seem to have higher SFE than old radio galaxies. 

Even though the detection of \hi\ and ionized gas outflows confirms the presence of AGN feedback in the host galaxies of young radio sources, like the ones examined in this work, the kinematics of the cold molecular gas and the SFE of the hosts seem, however,  mildly affected. 
It is possible that, due to the extreme youth of the AGN, the effects of feedback in young sources have not had the time to propagate to such large scales as in the evolved radio sources. 

\section{Summary and conclusions}

We have used the IRAM PdBI to study the distribution and kinematics of molecular gas {of the nearby  ($z$$\sim$0.1) FR~II radio galaxy \tcd}, by imaging  with high spatial 
resolution (0.6$\arcsec$) the emission of the 2--1 line of $^{12}$CO in the circumnuclear region of the galaxy. AGN activity has 
been recently ($\sim$10$^5$ yr ago) re-activated in \tcd\ , triggered by a merger episode ({that} occurred $\sim$10$^7$yr ago).  Previous observations identified in \tcd\ one of the most extreme \hi\ outflows thus far discovered in a radio galaxy.  The new observations presented in this paper have been analyzed to search for the footprints of AGN 
feedback on the molecular ISM of  \tcd. In particular, we have looked for evidence of outflow motions in the molecular gas kinematics. Furthermore, we have derived the SFE of 
molecular gas in \tcd\ and compared this value with the SFE obtained in different populations of radio galaxies, including $young$ and $evolved$ radio sources. We have investigated
if the star-formation properties of \tcd\ deviate from the KS relation followed by normal star-forming galaxies.  

We summarize below the main results and conclusions of this work:

\begin{itemize}

\item

The {CO  emission} comes from a spatially resolved 2.6~kpc-diameter disk that has a gas mass  \mhd=2.1$\times$10$^9$ \msun. The CO  disk is  linked to the inner
region of a highly-inclined dusty disk identified in the V--H HST color image of the galaxy. The molecular disk is fueling a star 
formation episode singled out by a chain of strong emission knots detected at UV wavelengths in the HST pictures of  \tcd\ .

\item

The kinematics of the CO disk are dominated by circular rotation and show no indications of high-velocities attributable to outflows. However, based on the limits imposed by the sensitivity and velocity coverage of our data, we cannot exclude in \tcd\ the existence of a molecular gas outflow {comparable} to the one detected in \object{Mrk~231}.
The gas disk rotates regularly around the AGN core. The latter is 
identified by prominent radio continuum emission detected at 3~mm and 1~mm wavelengths in our map.   Based on the observed CO kinematics we determine an upper limit for the redshift of the 
source $z_{\rm{CO}}$=0.09927$\pm$0.0002 (\vsco=29\,761$\pm$40~km s\mone).  The new value of \vs\  is significantly blue-shifted ($>$350~kms$^{-1}$) with respect to  the value of the 
systemic velocity previously reported in the literature for the source. 

\item

In the light of the new redshift value, the bulk of the deep  \hi\  absorption can be simply explained by a rotating \hi\ structure, 
leaving out the evidence of outflow only to the most extreme velocities (--1000~km s\mone$<$$\varv-$\vsco$<$--500~km s\mone). As for the CO emitting gas, outflow signatures are also absent in the warm 
molecular gas emission traced by infrared \hd\ lines.  A reanalysis of the ionized gas kinematics reveals nevertheless the existence of a fast outflow visible in the high-velocity  ($\sim$1000~km s\mone) 
blue and red-shifted line emission wings.

\item

 We have derived the SFR of \tcd\ using the different tracers available at optical, UV and MIR wavelengths. The most reliable SFR value comes from high-resolution UV-optical photometry, which in \tcd\ seems mostly unaffected by AGN contribution. Most of the other SFR tracers examined may have contributions from the AGN, however. The consistency of the SFR values obtained with all SFR tracers suggests nevertheless that the AGN contribution is kept low. High-resolution spectroscopy of the UV star-forming knots would be needed to accurately measure the AGN contribution.

\item

The SFE of the molecular disk, defined as $SFE$=$SFR/M_{gas}$,  
is  fully consistent with the value measured in $normal$ galaxies, i.e., systems that form 
stars following the canonical KS law. This result is at odds with the factor of 10--50 lower values that have been claimed  to characterize the $SFE$ of $evolved$ powerful radio galaxies. As an 
explanation for the different SFE measured in  $young$ and $evolved$ radio sources, the most likely scenario in \tcd\ suggests that none of the effects of AGN feedback have yet had the time to affect 
the kinematics or the SF properties of the molecular ISM due to the extreme youth of the AGN re-activation episode in this source.

\end{itemize}

This simple evolutionary scenario remains to be confirmed using larger samples, and most notably, using different tracers of the SFR in radio galaxies.
As a caveat, it should be pointed out that most of the estimates of the SFR in $evolved$ radio sources have been thus far derived from the strength of the PAH bands at 7.7 $\mu$m.
The $low$ SFR values in these sources may be severely underestimated due to the destruction of PAH carriers  by shocks and/or strong UV and X-ray  fields; these ingredients are 
known to be in place particularly in radio-loud AGN. The reported differences in the $SFE$ of $young$ and $evolved$
radio galaxies could simply reflect a mismatch in the calibration of the SFR tracers used.   






\begin{acknowledgements}

We are grateful to Dr. C. Struve, Dr. J. E. Conway and Prof. M. Birkinshaw for sharing their results before publication, and to Dr. D. Dicken for useful discussions.  AL acknowledges support by the Spanish MICINN within the program CONSOLIDER INGENIO 2010, under grant ASTROMOL (CSD2009-00038), Springer and EAS. This research has made use of NASA's Astrophysics Data System Bibliographic Services and of the NASA/IPAC Extragalactic Database (NED) which is operated by the Jet Propulsion Laboratory, California Institute of Technology, under contract with the National Aeronautics and Space Administration. Based on observations made with the NASA/ESA Hubble Space Telescope, and obtained from the Hubble Legacy Archive, which is a collaboration between the Space Telescope Science Institute (STScI/NASA), the Space Telescope European Coordinating Facility (ST-ECF/ESA) and the Canadian Astronomy Data Centre (CADC/NRC/CSA). Funding for the SDSS and SDSS-II has been provided by the Alfred P. Sloan Foundation, the Participating Institutions, the National Science Foundation, the United States Department of Energy, NASA, the Japanese Monbukagakusho, the Max-Planck Society, and the Higher Education Funding Council for England. The SDSS website is http://www.sdss.org/. The SDSS is managed by the Astrophysical Research Consortium for the Participating Institutions. The Participating Institutions are the American Museum of Natural History, Astrophysical Institute Potsdam, University of Basel, University of Cambridge, Case Western Reserve University, University of Chicago, Drexel University, Fermilab, the Institute for Advanced Study, the Japan Participation Group, Johns Hopkins University, the Joint Institute for Nuclear Astrophysics, the Kavli Institute for Particle Astrophysics and Cosmology, the Korean Scientist Group, the Chinese Academy of Sciences (LAMOST), Los Alamos National Laboratory, the Max-Planck Institute for Astronomy (MPIA), the Max-Planck Institute for Astrophysics (MPA), New Mexico State University, Ohio State University, University of Pittsburgh, University of Portsmouth, Princeton University, the United States Naval Observatory and the University of Washington. This work makes use of euro-vo software, tools or services and TOPCAT \citep{Taylor05}. Euro-vo has been funded by the European Commission through contract numbers RI031675 (DCA) and 011892 (VO-TECH) under the Sixth Framework Programme and contract number 212104 (AIDA) under the Seventh Framework Programme.

\end{acknowledgements}

\bibliographystyle{aa}
\bibliography{../../../ALORefs}

\appendix
\section{Spatially-resolved SFE of \tcd}
\label{appregions}

\begin{figure}[b]
\centering
\includegraphics[width=\columnwidth]{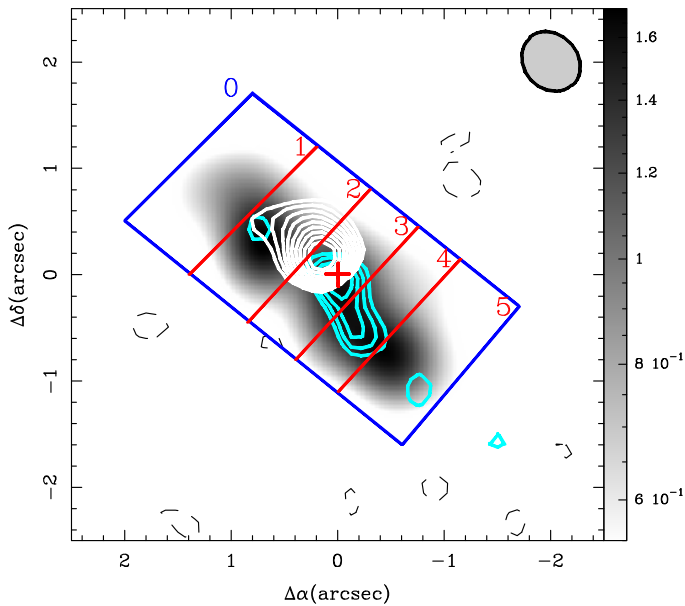} 
\caption{HST FUV image of the star forming regions of \tcd, smoothed with the beam of CO, with the red and blue \codu\ channels overlaid. Contour levels as in Fig. \ref{vhoverlay}. Rectangles show the schematic location of the different regions for the analysis of the SFR and \hd\ distribution. Color version available in electronic format.  \label{regions}}
\end{figure}


The emission of the star forming regions and the CO of \tcd\ are almost co-spatial, suggesting that \tcd\ is using the cold molecular gas traced by CO to form stars. The resolution of the HST images and the CO map are enough to study the SFE along the molecular gas disk of \tcd. Therefore, we  searched for the star-forming region responsible of the  normal (KS-law) SFE of \tcd.

Firstly, we smoothed the HST FUV  image with the beam of the \codu\ maps, to have the same resolution in both images. We then divided the disk in five regions, based on the UV emission and molecular gas content.  
Figure \ref{regions} shows the smoothed FUV image with the \codu\ blue and red channels overlaid, and the location of the five regions. 
Regions 1 and 5 correspond to the most external UV emission 
with no \codu\ detection above 3$\sigma$ levels. Regions 2 and 4 have UV 
and \codu\ emission. Region 3 corresponds to the center of the disk. We named ``Region 0'' the sum of regions 1 through 5. 

$SFR$=7.51 \msun\ yr\mone\ was calculated using all flux down to $1\sigma$ over the background \citep[aperture ``All'',][]{Tremblay10}. 
To obtain the local SFR for each region, we divided 7.51 \msun\ yr\mone\ by the flux in the ``All'' aperture, and multiplied by the flux of each region. The \hd\ masses were measured for each region individually, using the \codu\ emission map. The SFR and \hd\ contents for each region are listed in Table \ref{tab_sfr}. 
The SFE of the  star forming regions are $SFE_{\rm{R1}}$$>$0.8 Gyr\mone,  $SFE_{\rm{R2}}$=3 Gyr\mone,  $SFE_{\rm{R3}}$=3 Gyr\mone,  $SFE_{\rm{R4}}$=4 Gyr\mone, and  $SFE_{\rm{R5}}$$>$0.9 Gyr\mone, consistent with 
 the SFE calculated using the ``All'' aperture, and 
the SFE of galaxies following the KS-law \citep[e.g., SINGS sample,][]{Kennicutt03, Roussel07}. The SFE of \tcd\ is not a local effect from one individual region, \tcd\ shows  normal SFE all along its molecular gas disk.

\end{document}